\documentstyle[aps, epsf, amsmath, fleqn]{revtex}

\def \half{\frac{1}{2}}
\def \eps{\epsilon}
\def \d{\partial}
\def \lam{\lambda}

\renewcommand{\(}{\left(}
\renewcommand{\)}{\right)}
\newcommand{\df}[2]{ \frac{\partial {#1}}{\partial {#2}} }
\newcommand{\dff}[2]{ \frac{\partial^2 {#1}}{\partial {#2}^2} }
\newcommand{\wh}{\widehat}

\begin{document}

\title{New approach to the evolution of neutron star oscillations}

\author{Johannes Ruoff}

\address{Institut f\"ur Astronomie und Astrophysik\\
Universit\"at T\"ubingen, D-72076 T\"ubingen, Germany}

\maketitle

\begin{abstract}
  We present a new derivation of the perturbation equations governing
  the oscillations of relativistic non-rotating neutron star models
  using the ADM-formalism. This formulation has the advantage that it
  immediately yields the evolution equations in a hyperbolic form,
  which is not the case for the Einstein field equations in their
  original form. We show that the perturbation equations can always be
  written in terms of spacetime variables only, regardless of any
  particular gauge. We demonstrate how to obtain the Regge-Wheeler
  gauge, by choosing appropriate lapse and shift. In addition, not
  only the 3-metric but also the extrinsic curvature of the initial
  slice have to satisfy certain conditions in order to preserve the
  Regge-Wheeler gauge throughout the evolution.  We discuss various
  forms of the equations and show their relation to the formulation of
  Allen et al.  New results are presented for polytropic equations of
  state. An interesting phenomenon occurs in very compact stars, where
  the first ring-down phase in the wave signal corresponds to the
  first quasinormal mode of an equal mass black hole, rather than to
  one of the proper quasinormal modes of the stellar model. A somewhat
  heuristic explanation to account for this phenomenon is given. For
  realistic equations of state, the numerical evolutions exhibit an
  instability, which does not occur for polytropic equations of state.
  We show that this instability is related to the behavior of the
  sound speed at the neutron drip point. As a remedy, we devise a
  transformation of the radial coordinate $r$ inside the star, which
  removes this instability and yields stable evolutions for any chosen
  numerical resolution.
\end{abstract}


\widetext

\section{Introduction}
\label{sec:introduction}
The driving force behind the theoretical studies in gravitational
radiation is to predict and investigate the sources of gravitational
waves that will be observed by interferometric (GEO600, LIGO, VIRGO,
TAMA,LISA) and bar (ALLEGRO, AURIGA, EXPLORER, NAUTILUS) detectors. The
goal of these studies is not limited to facilitating the detection of
signals but also to provide tools to extract as much information as
possible about the physical nature of the sources. Neutron stars have
been pointed out as obvious source candidates, since one channel by
which an oscillating neutron star looses energy is via the emission of
gravitational radiation.

This radiation basically consists of a superposition of its
characteristic oscillatory modes, which can be grouped into two
families: i) The fluid modes \cite{Th69a,LD83}, which have a Newtonian
counterpart, but which through their coupling to the spacetime are
damped because they now can emit gravitational waves. ii) The
spacetime modes \cite{Koj88,KS92,LNS93}, which have no Newtonian
counterpart, and which couple only weakly to the fluid. (In the odd
parity case, they do not couple to the fluid at all). These modes
usually are strongly damped, however, for ultra-relativistic stars,
the spacetime curvature can be so strong that it can trap impinging
gravitational waves. Those ``trapped'' modes \cite{CF91b,Kok94} are
quite long lived, since they only leak out slowly from inside the
gravitational potential well created by the neutron star. For recent
comprehensive reviews on oscillation modes of neutron stars and black
holes, see \cite{KS99,Nollert99}.

However, just knowing the different oscillation modes of a neutron
star is not enough. To be able to detect and to interpret a signal in
a gravitational wave detector, it is crucial to have accurate
templates of the waveforms that result from astrophysical events, such
as the formation of the neutron star after the collapse of a
progenitor star at the end point of nuclear burning. To that purpose,
two ingredients are necessary. Firstly, one has to have initial data
that represent a realistic stage of the astrophysical system under
consideration. Secondly, one needs a numerical evolution code that can
evolve these initial data in a stable and accurate way.

It is only by fully nonlinear simulations that accurate waveforms of
oscillating neutron stars after core collapse or binary merger can be
obtained. Therefore it is a major goal of several groups to push the
further development of nonlinear evolution codes. The biggest
obstacle, however, that still prevents one from obtaining accurate
results, is the enormous computational expenditure to solve the full
set of Einstein equations in 3D. Even modest resolutions easily exceed
the capabilities of the largest and fastest present day computers.

It is therefore crucial to have some alternative methods which require
less computing power but still give valuable physical insight. One of
them is perturbation theory. In the non-rotating case, due to the
spherical background, one can completely separate off the angular
dependence, and the evolution problem boils down to the integration of 1D
wave equations. Slowly rotating stars can still be treated in 1D as long
as the deformation due to rotation is negligible. But for rapidly
rotating neutron stars, the spherical symmetry is broken, and one can
only separate off the azimuthal angular dependence. Hence the problem
becomes 2D, which is nevertheless still numerically tractable. The
results from those linear evolutions are expected to give accurate
waveforms in the range where the neutron star oscillates only weakly.
Furthermore, they can be used as testbeds for the nonlinear
calculations. Indeed, recent results of nonlinear evolutions have
confirmed the existence of the various modes predicted by linear
perturbation theory \cite{FSK99}.

The evolution of polar perturbations of non-rotating stars has been
first addressed by Allen et al.~\cite{AAKS98}. In their formulation
they describe the oscillations with three coupled wave equations, two
for the metric perturbations in and outside the star, and one for the
fluid inside the star. The evolution of various initial data have
shown that generic initial data can indeed excite some of the stellar
modes such as the $f$-mode, some $p$-modes and the first $w$-mode.

In this paper we rederive the perturbation equations using the (3+1)
or ADM-form \cite{ADM} of the Einstein equations. These equations are
much better suited for the numerical initial value problem, since they
immediately yield a hyperbolic evolution system provided that a
suitable gauge has been chosen, whereas the original field equations
contain mixed time and space derivatives. It is only by the
introduction of new variables that those equations can be cast into a
well suited form. The ADM-formalism has already been addressed by
Moncrief \cite{Mon74a,Mon74b}, where it was used to derive a
Hamiltonian gauge invariant formulation of the perturbation equations.
Albeit preferable from a conceptual point of view, since they are
independent of any gauge, those equations do not prove particularly
useful for numerical evolutions. Our starting point therefore are the
field equations written as a set of evolution equations for the metric
and the extrinsic curvature of a 3-dimensional spatial hypersurface,
together with the constraint equations, which have to be satisfied at
every instance of time.

As an immediate consequence of using the ADM-formalism, it follows
that it is always possible to eliminate the fluid variables and thus
to write down the evolution equations in terms of metric quantities
only. This has already been shown by Chandrasekhar and Ferrari
\cite{CF91a} for the diagonal gauge and by Ipser and Price \cite{IP91}
for the Regge-Wheeler gauge. However, from our formalism it is clear
that this feature is independent of the chosen gauge.

After having expanded the equations in spherical harmonics, we will
choose the Regge-Wheeler gauge as the basis of further investigations.
Due to the spherical harmonics the equations are independent of the
azimuthal order $m$ and can be divided into two uncoupled sets
according to their behavior under parity transformation. The polar or
even parity equations transform as $(-1)^l$, whereas the axial or odd
parity equations as $(-1)^{l+1}$. Our main focus will be the polar
perturbations, nevertheless, due to their simplicity, we also present
the axial equations in the ADM-form. The derivation of the polar
equations is much more involved, because the ``raw'' forms lead to
numerical instabilities at the origin due to indefinite expressions,
which is a well known consequence of using spherical coordinates. We
are therefore forced to recast the polar equations into a form that is
well behaved at the origin.

Having forged the equations into a form that yields stable evolutions
for polytropic equations of state, we face a new instability close to
the stellar surface when we try to switch to realistic equations of
state. In this paper we use a quite recent equation of state, called
MPA \cite{Wu91}. Yet, this instability is not a consequence of just
this particular equation of state, it will occur for any realistic
equation of state. As we shall show, the instability results from the
behavior of the sound speed at the neutron drip point, where the
equation of state suddenly becomes very soft. This is accompanied by a
sharp drop of the sound speed, which can give rise to a numerical
instability if the spatial grid resolution is chosen too low. However,
by increasing the resolution this instability will eventually vanish.
In the non-rotating case, having to resort to higher resolutions does
not present too great an obstacle, since the equations are 1D wave
equations. However, this instability presumably also occurs for
rotating stars, where the equations are 2D, and the required
resolution in order to obtain stable evolutions might stretch the
computing times to unacceptably large values. Therefore we devise a
coordinate transformation of the radial coordinate inside the star,
which enables stable evolutions for any spatial resolution.
 
The remainder of this paper is organized as follows: In Sec.~II we
derive the relevant equations. Section III is devoted to a discussion
of the boundary conditions. In Sec.~IV we briefly comment on the
construction of initial data, and in Sec.~V we present results for
polytropic equations of state. Finally, in Sec.~VI we discuss the
instability associated with the use of realistic equations of state,
and we present the coordinate transformation that removes it.
Conclusions are briefly presented in Sec.~VII.

We adopt the metric signature ($-$+++) and work in geometric units
with $G = c = 1$. Spacetime and spatial indices are denoted by Greek
and Latin letters, respectively. Derivatives with respect to the
radial coordinate $r$ are sometimes denoted by a prime and derivatives
with respect to the time coordinate $t$ by an overdot.

\section{Derivation of the perturbation equations}

\subsection{The unperturbed stellar model}
\label{sec:model}
The background geometry of a non-rotating, spherically symmetric
star is given by the line element 
\begin{align}
  ds^2_{(0)} &= -e^{2\nu} dt^2 + e^{2\lambda} dr^2 + r^2(d\theta^2
  + \sin^2\theta d\phi^2)\;,
\label{bg}
\end{align}
where $\nu$ and $\lambda$ are functions of the radial coordinate $r$.
We model the star as a perfect fluid whose energy-momentum tensor has
the form
\begin{align}\label{Tmunu}
  T_{\mu\nu} &= (p +\eps)\,u_\mu u_\nu + p\,g_{\mu\nu}\;,
\end{align}
with $p$ denoting the pressure, $\eps$ the energy density, and $u^\mu$
the 4-velocity of the fluid. In the fluid rest frame $u^\mu$ has only
one non-vanishing component $u^0 = e^{-\nu}$. Einstein's equations
$G_{\mu\nu} = 8\pi T_{\mu\nu}$ and the conservation equations
$T^{\mu\nu}_{\phantom{\mu\nu};\nu} = 0$ yield the following three
independent structure equations for the four unknown $\lambda,\,
\nu,\, p$ and $\eps$:
\begin{subequations}
\label{toveqns}
\begin{align}
  \label{tov1}\lam' &= \frac{1 - e^{2\lam}}{2r}
  + 4\pi r e^{2\lam}\epsilon \\
  \label{tov2}\nu' &= \frac{e^{2\lam} - 1}{2r}
  + 4\pi r e^{2\lam}p\\
  \label{tov3}p' &= - \nu'(p + \epsilon)\;.
\end{align}
\end{subequations}
We will refer to those equations as Tolman-Oppenheimer-Volkoff (TOV)
equations, although they are usually written in terms of the
gravitational mass function $m$, which is related to $\lam$ through
\begin{align}
  e^{-2\lam} \;&\equiv\; 1 - \frac{2m}{r}\;.
\end{align}
To fully determine this system of equations, an equation of state must
be supplemented. In this paper, we restrict ourselves to barotropic
equations of state, where the pressure is a function of the energy
density alone:
\begin{align}\label{eos}
p &= p(\eps)\;.
\end{align}
In particular, we will use either a polytropic equation of state in
form of
\begin{align}
  p &= \kappa\eps^\Gamma
\end{align}
or a tabulated realistic zero temperature equation of state. To obtain
the stellar model, we have to integrate the TOV equations
\eqref{toveqns} together with the equation of state \eqref{eos} from
the center up to the point where the pressure $p$ vanishes. This then
defines the surface $R$ of the star.  From Birkhoff theorem it follows
that the exterior vacuum region of the star is described by the
Schwarzschild metric with the mass parameter $M \equiv m(R)$.
\subsection{The perturbation equations}

\subsubsection{The general form}

Let us recall that the general line element, when written
in the ADM-form, reads
\begin{align}\label{ADMmetric}
  ds^2 &= -\(\alpha^2 - \beta_k\beta^k\)dt^2 + 2\beta_i\,dt\,dx^i
  + \gamma_{ij}\,dx^idx^j\;,
\end{align}
where $\alpha$, $\beta^k$, and $\gamma_{ij}$ are lapse function, shift
vector, and spatial metric, respectively. Comparing the ADM-metric
\eqref{ADMmetric} with the background metric \eqref{bg} reveals that
the background shift $\beta^k_{(0)}$ vanishes and the background lapse
function is $\alpha_{(0)} = e^\nu$. In addition, the extrinsic
curvature of constant $t$ slices vanishes, since the background metric
is static.

Let $\alpha$ and $\beta_i$ now be the perturbations of the lapse
function and the (covariant) shift vector, and let $h_{ij}$ denote the
perturbations of the spatial metric. The line element describing the
perturbations to the background metric \eqref{bg} then reads
\begin{align}\label{eq:perturb}
  ds^2_{(1)} &= -2e^\nu\alpha\,dt^2 + 2\beta_i\,dt\,dx^i + h_{ij}\,dx^idx^j\;.
\end{align}
The perturbations of the energy-momentum tensor \eqref{Tmunu} can be
written in terms of $\delta\eps$, $\delta p$, and $\delta u_\mu$,
which are the (Eulerian) perturbations of the energy density $\eps$,
pressure $p$, and (covariant) 4-velocity $u_\mu$, respectively. We
assume the equation of state \eqref{eos} to hold for both the
unperturbed and perturbed configurations (isentropic perturbations),
therefore the perturbations $\delta\eps$ and $\delta p$ are related
through
\begin{align}
  \delta p &= \frac{dp}{d\eps}\,\delta\eps
  = \frac{p'}{\eps'}\,\delta\eps
  = C_s^2\delta\eps\;,
\end{align}
where $C_s$ is the sound speed inside the fluid.

The dynamic equations that govern the metric and extrinsic curvature
perturbations $h_{ij}$ and $k_{ij}$ read after linearization
\begin{align}\label{hij}
  \d_t h_{ij} &= \beta^k\d_k\gamma_{ij} + \gamma_{ki}
  \d_j\beta^k + \gamma_{kj} \d_i\beta^k
  - 2 e^\nu k_{ij}\\
  \begin{split}\label{kij}
    \d_t k_{ij} &= - \d_i \d_j \alpha +
    \Gamma ^k_{\phantom{i}ij}\d_k\alpha + \delta \Gamma
    ^k_{\phantom{i}ij}\d _k e^\nu
    + \alpha\left[R_{ij} + 4\pi \(p - \eps\)\gamma_{ij}\right]\\
    &\quad{} + e^\nu\left[\,\delta R_{ij} + 4\pi 
      \(\(p - \eps\)h_{ij} + \delta\eps\(C^2_s -
      1\)\gamma_{ij}\) \right]\;.
  \end{split}
\end{align}
Therein, the $\delta\Gamma^k_{\phantom{i}ij}$ represent the
perturbations of the spatial Christoffel symbols
$\Gamma^k_{\phantom{i}ij}$
\begin{align}
\begin{split}
  \delta\Gamma^k_{\phantom{i}ij} &= \half \gamma^{kl}
  \( h_{li;j} + h_{lj;i} - h_{ij;l}\)\\
  &= \half \gamma^{kl}
  \(\d_j h_{li} + \d_i h_{lj} - \d_l h_{ij}\)
  - h_{lm}\gamma^{kl}\Gamma^m_{ij}
\end{split}
\end{align}
and $\delta R_{ij}$ the perturbations of the spatial Ricci tensor
$R_{ij}$
\begin{align}
\begin{split}
  \delta R_{ij} &=  \delta\Gamma^k_{\phantom{i}ij;k}
  - \delta\Gamma^k_{\phantom{i}ik;j}\\
  &= \d_k\delta\Gamma^k_{\phantom{i}ij}
  - \d_j\delta\Gamma^k_{\phantom{i}ik}
  - \delta\Gamma^k_{\phantom{i}mk}\Gamma^m_{\phantom{i}ij}
  + \delta\Gamma^m_{\phantom{i}ij}\Gamma^k_{\phantom{i}mk}
  - \delta\Gamma^k_{\phantom{i}mj}\Gamma^m_{\phantom{i}ik}
  - \delta\Gamma^m_{\phantom{i}ik}\Gamma^k_{\phantom{i}mj}\;.
\end{split}
\end{align}
We should note that the indices of perturbations are lowered or raised
with the background 3-metric $\gamma_{ij}$. In constructing initial
data, the perturbations must satisfy the linearized version of
Einstein's constraint equations, namely
\begin{align}\label{hhc}
  \gamma^{ij}\delta R_{ij} - h^{ij}R_{ij} &= 16 \pi \delta \eps\\
  \gamma^{jk}\(\d _i k_{jk} - \d _j k_{ik}
  - \Gamma^l_{\phantom{i}ik} k_{jl}
  + \Gamma^l_{\phantom{i}jk} k_{il}\)
  &= -8\pi(p + \eps)\delta u_i\,.\label{mmc}
\end{align}
From the linearized energy-momentum conservation equations $\delta
T^{\mu\nu}_{\phantom{\mu\nu};\nu} = 0$, we can deduce equations of
motion for the fluid perturbations $\delta \eps$ and $\delta u_i$,
which complete the set of dynamical equations.

However, we can as well dispense with the fluid equations, since it is
possible to eliminate the energy density perturbation $\delta\eps$
from the evolution equation \eqref{kij} by virtue of the Hamiltonian
constraint \eqref{hhc}. In this way we can obtain a consistent system
of evolution equations for the metric and extrinsic curvature
perturbations alone. The constraints then can serve as a means to
compute the matter perturbations $\delta\eps$ and $\delta u_i$. We
should stress that the possibility to completely describe the
oscillations of neutron stars with spacetime variables only is a
general feature of the perturbation equations and does not depend on
any specific gauge choice.

Due to the spherical symmetry of the background, we can eliminate the
angular dependence of the perturbation equations by expanding them
into spherical tensor harmonics. Those tensor harmonics can be divided
into two subsets that behave differently under parity transformation.
Under space reflection the {\it even parity} or {\it polar} harmonics
change sign according to $(-1)^l$, whereas the {\it odd parity} or
{\it axial} harmonics transform like $(-1)^{l+1}$. Here, $l$ is the
number that labels the multipole of the spherical harmonics
$Y_{lm}$. Because of the spherical symmetry, the axial and polar
equations are decoupled and degenerate with respect to the 
azimuthal number $m$.

Without picking a specific gauge, the metric perturbations can be
expanded using the tensor harmonics given first by Regge and Wheeler
\cite{RW57} in the following way (symmetric components are denoted by
an asterisk):
\begin{subequations}
  \begin{align}\label{exp_h}
    \alpha &= \sum_{l = 2}^{\infty} \sum_{m = -l}^le^\nu\wh{S}_1^{lm}Y_{lm}\\
    \beta_k &= \sum_{l = 2}^{\infty} \sum_{m = -l}^l\(\wh{S}_2^{lm}Y_{lm}, \wh{V}_1^{lm}\Psi^{lm}_\alpha
    + \wh{V}_2^{lm}\Phi^{lm}_\alpha\)\\
    h_{ij} &= \sum_{l = 2}^{\infty} \sum_{m = -l}^l
    \( \begin{array}{ccc}
      e^{2\lam}\wh{S}_3^{lm}Y_{lm} &
      \wh{V}_3^{lm}\Psi^{lm}_\alpha
      + \wh{V}_4^{lm}\Phi^{lm}_\alpha\\
      {}* &
      r^2\(\wh{T}_1^{lm}\Psi^{lm}_{\alpha\beta}
      + \wh{T}_2^{lm}\Phi^{lm}_{\alpha\beta}
      + \wh{T}_3^{lm}\chi^{lm}_{\alpha\beta}\)\\
    \end{array} \) \;,
  \end{align}
\end{subequations}
where
\begin{align}
  \Psi^{lm}_\alpha &= \(\df{}{\theta},\;\df{}{\phi}\)Y_{lm}\;,\\
  \Phi^{lm}_\alpha &= \(-\sin^{-1}\theta\df{}{\phi},\;
  \sin\theta\df{}{\theta}\)Y_{lm}\;,
\end{align}
and
\begin{align}
  \Phi^{lm}_{\alpha\beta} &=  \( \begin{array}{cc}
    1 & 0 \\
    0 & \sin^2\theta
  \end{array} \)Y_{lm}\\
  \Psi^{lm}_{\alpha\beta} &= \( \begin{array}{cc}
    W_{lm} & X_{lm} \\
    X_{lm} & - \sin^2\theta W_{lm}
  \end{array}\) - \half l(l+1)\Phi^{lm}_{\alpha\beta} \\
  \chi^{lm}_{\alpha\beta} &=  \sin\theta \( \begin{array}{cc}
    -\sin^{-2}\theta X_{lm} & W_{lm} \\
    W_{lm} & X_{lm}
  \end{array} \)\;,
\end{align}
with
\begin{align}
  \begin{split}
    W_{lm} &= \half\(\dff{}{\theta} - \cot\theta\df{}{\theta}
    - \sin^{-2}\theta\dff{}{\phi}\)Y_{lm}\\
    &= \(\half l(l+1) + \dff{}{\theta}\)Y_{lm}
  \end{split}\\
  X_{lm} &= \(\df{}{\theta} - \cot\theta\)\df{}{\phi}Y_{lm}\;.
\end{align}
The notation has been chosen such that the coefficients $\wh{S}_i$
represent the {\em scalar} parts of $h_{\mu\nu}$, namely $\alpha,
\beta_r$, and $h_{rr}$, whereas the $\wh{V}_i$ stand for the {\em
vector} components $\beta_\theta, \beta_\phi, h_{r\theta}$ and
$h_{r\phi}$. Lastly, the $\wh{T}_i$ represent the {\em tensor}
components $h_{\theta\theta}, h_{\theta\phi}$ and $h_{\phi\phi}$. Note
that this expansion includes both polar and axial harmonics. The
latter are represented by $\Phi^{lm}_\alpha$ and
$\chi^{lm}_{\alpha\beta}$ with respective coefficients $\wh{V}_2,
\wh{V}_4$ and $\wh{T}_3$. Similarly, the extrinsic curvature tensor
$k_{ij}$ will be expanded as
\begin{align}\label{exp_k}
  k_{ij} &= \sum_{l = 2}^{\infty} \sum_{m = -l}^l
  \( \begin{array}{ccc}
    e^{2\lam}\wh{K}_1^{lm}Y_{lm} &
    \wh{K}_2^{lm}\Psi^{lm}_\alpha
    + \wh{K}_3^{lm}\Phi^{lm}_\alpha\\
    {}* &
    r^2\(\wh{K}_4^{lm}\Psi^{lm}_{\alpha\beta}
    + \wh{K}_5^{lm}\Phi^{lm}_{\alpha\beta}
    + \wh{K}_6^{lm}\chi^{lm}_{\alpha\beta}\)\\
  \end{array} \) \;.
\end{align}
Here, $\wh{K}_3^{lm}$ and $\wh{K}_6^{lm}$ are the axial coefficients.

Last not least, we need the matter variables
\begin{align}
  \delta\eps &= \sum_{l = 2}^{\infty} \sum_{m = -l}^l
  \hat{\rho}^{lm}Y_{lm}\label{delta_eps}\\
  u_i &= \sum_{l = 2}^{\infty} \sum_{m = -l}^l\(\hat{u}_1^{lm}Y_{lm},\,
  \hat{u}_2^{lm}\Psi^{lm}_\alpha + 
  \hat{u}_3^{lm}\Phi^{lm}_\alpha\)\label{u_i}\;.
\end{align}
The sum over the multipoles starts from $l=2$, since we are only
interested in perturbations which are associated with the emission of
gravitational radiation. Besides, the non-radiative multipoles $l=0$
and $l=1$ would have to be treated in a different way.

Using the above expansions, we obtain 12 evolution equations for the
coefficients of the 3-metric $h_{ij}$ and the extrinsic curvature
$k_{ij}$. However, it is clear that this set cannot be used for the
evolution, for we have not specified any gauge yet. Within
perturbation theory, picking a specific gauge usually means to define a
gauge vector $\eta^\mu$ and transform the metric perturbations
according to
\begin{align}
  h^{new}_{\mu\nu} &= h^{old}_{\mu\nu}
  - \eta_{\mu;\nu} - \eta_{\nu;\mu}\:.
\end{align}
In our case, however, we follow the spirit of the ADM-formalism, that
is we pick the gauge by prescribing the coefficients of lapse $\alpha$
and shift $\beta_i$, for which we do not have any evolution equations.
Yet, this is not enough as we shall see. To fully fix the gauge, we
also have to impose certain constraints on the initial data.

The most common gauge is the Regge-Wheeler gauge \cite{RW57}, which
can be obtained by setting a certain number of metric perturbation
coefficients to zero. Translating this into the ADM-formalism means to
choose shift and lapse in such a way that for initial data which obey
the Regge-Wheeler gauge, the evolution will preserve it.

Regge and Wheeler have chosen their gauge such that the metric
coefficients $\wh{V}_1^{lm}, \wh{V}_3^{lm},\wh{T}_1^{lm}$, and
$\wh{T}_3^{lm}$ vanish. By choosing shift and lapse as (from now on we
omit the indices $l$ and $m$)
\begin{align}
  \wh{S}_1 &= -\half\wh{S}_3\\
  \wh{S}_2 &= 2e^\nu \wh{K}_2\\
  \wh{V}_1 &= 0\\ \wh{V}_2 &= e^\nu \wh{K}_6\;,
\end{align}
we obtain the following evolution equations for the coefficients
$\wh{V}_3,\wh{T}_1$ and $\wh{T}_3$:
\begin{align}
  \df{}{t}\wh{V}_3 &= 0\\
  \df{}{t}\wh{T}_1 &= -2e^\nu \wh{K}_4\label{T1}\\
  \df{}{t}\wh{T}_3 &= 0\;.
\end{align}
In addition the evolution equation for the extrinsic curvature
component $\wh{K}_4$ depends only on $\wh{T}_1$ and $\wh{V}_3$. Thus,
only for initial data satisfying $\wh{V}_3 = \wh{T}_1 = \wh{T}_3 = 0$
{\em and} $\wh{K}_4 = 0$, the evolution equations guarantee the vanishing of
those coefficients throughout the whole evolution.

Hence, the Regge-Wheeler gauge does not only impose constraints on the
initial metric perturbations but also on the extrinsic curvature
perturbations. The vanishing of the coefficient $\wh{K}_4$ is a
crucial feature of the Regge-Wheeler gauge and has to be imposed on
the initial data. This is in contradiction with the statement of
Andersson et al.~\cite{AKLPS99}, where the authors claim that there
are no constraints on the odd or even parity nature of the extrinsic
curvature. This is not true, since the vanishing of $\wh{K}_4$ {\em
  must} be ensured on the initial slice, otherwise the evolution {\em
  cannot} be performed in the Regge-Wheeler gauge. For data with
non-vanishing $\wh{K}_4$, we would have $\d\wh{T}_1/\d t \ne 0$, which is
incompatible with the Regge-Wheeler gauge, because it would lead to a
non-vanishing $\wh{T}_1$ even if $\wh{T}_1$ was initially set to zero.
Therefore the initial data presented in \cite{AKLPS99} for two
colliding neutron stars are not adequate for an evolution code which
makes use of the Regge-Wheeler gauge, since the coefficient $k_4$ in
Eqs.~(82) of \cite{AKLPS99} (which corresponds to our $\wh{K}_4$)
does not vanish. We will discuss the issue of constructing initial
data in more detail in section \ref{sec:init}.

Having switched to the Regge-Wheeler gauge, we are left with evolution
equations for the three metric variables $\wh{S}_3$, $\wh{V}_4$, and
$\wh{T}_2$, and all the extrinsic curvature variables $\wh{K}_i$ save
$\wh{K}_4$.  (In the more common notation of Regge and Wheeler
\cite{RW57}, it is $\wh{S}_3 = H_2$, $\wh{V}_4 = h_1$ and $\wh{T}_2 =
K$). We now split the equations with respect to their behavior under
parity transformation. Let us first focus our attention to the axial
case.

\subsubsection{Axial perturbations}

If in the expansions \eqref{exp_h} and \eqref{exp_k} we use new
variables
\begin{subequations}
\begin{align}
  V_4 &= e^{\nu-\lam}{\wh V}_4\\
  K_3 &= 2 e^{-\lam}{\wh K}_3\\
  K_6 &= r^2 e^{\lam}{\wh K}_6\;,
\end{align}
\end{subequations}
we obtain the following set of evolution equations:
\begin{subequations}
\label{axial}
\begin{align}
  \label{V4} \df{V_4}{t} &= e^{2\nu-2\lam}\(\df{K_6}{r}\
  + \(\nu' - \lam' - \frac{2}{r}\)K_6 - e^{2\lam}K_3\)\\
  \df{K_3}{t} &= \frac{l(l+1) - 2}{r^2}\,V_4\\
  \label{K6} \df{K_6}{t} &= \df{V_4}{r}\;,
\end{align}
\end{subequations}
and one constraint equation:
\begin{align}\label{oddMC}
  \df{K_3}{r} + \frac{2}{r}K_3 - \frac{l(l+1) - 2}{r^2}K_6
  &= 16\pi e^{\lam}(p + \eps){\hat u}_3\;.
\end{align}
From the conservation law $\delta T^{\mu\nu}_{\phantom{\mu\nu};\nu} = 0$, it
follows that
\begin{align}
  \df{{\hat u}_3}{t} &= 0\;, 
\end{align}
which means that axial gravitational waves do not couple to the
stellar fluid. Of course, this does not mean that there cannot be any
axial modes at all, in fact, there exists a whole spectrum of axial
spacetime modes, which is quite similar the polar spectrum
\cite{AKK96}. It is only the fluid modes that are missing.

The three evolution equations \eqref{axial} can be combined to yield a
single wave equation for the metric perturbation $V_4$, or better for
the variable
\begin{align}
  Q \;&:=\; \frac{V_4}{r}\;,
\end{align}
which completely describes the dynamical evolution of axial
perturbations:
\begin{align}\label{Qtt}
  \dff{Q}{t} &= \frac{\d^2Q}{\d r^2_*}
  + e^{2\nu}\( 4\pi\(p - \eps\) + \frac{6m}{r^3}
  - \frac{l(l+1)}{r^2}\)Q\;.
\end{align}
Herein, $r_*$ denotes the tortoise coordinate defined by
\begin{align}\label{tort}
  \frac{dr}{dr_*} &= e^{\nu-\lam}\;.
\end{align}
In the exterior, where $p$ and $\eps$ vanish and $m = M$, Eq.~\eqref{Qtt}
reduces to the well known Regge-Wheeler equation \cite{RW57}.

\subsubsection{Polar perturbations}

For the polar case we have only two dynamical metric perturbations
${\wh S}_3$ and ${\wh T}_2$, since the lapse ${\wh S}_1$ is
proportional to ${\wh S}_3$, and the only non-vanishing component ${\wh
S}_2$ of the shift is proportional to ${\wh K}_2$. To obtain the
relevant equations, we again choose a somewhat different
expansion. For the metric we use
\begin{align}
  \alpha &= -\half e^{\nu}\(\frac{T}{r} + rS\)Y_{lm}\\
  \beta_i &= \(e^{2\lambda}K_2,\,0,\,0\)Y_{lm}\\
  h_{ij} &= \label{hij_polar}
  \(\begin{array}{ccc}
    e^{2\lambda}\(\frac{T}{r} + rS\) & 0 & 0\\
    0 & rT & 0\\
    0 & 0 & r\sin^2 T
  \end{array}\)Y_{lm}\;,
\end{align}
and for the extrinsic curvature
\begin{align}
  k_{ij} &= 
  -\frac{e^{-\nu}}{2r}\(\begin{array}{ccc}
    e^{2\lambda}K_1 & -re^{2\lambda}K_2\df{}{\theta}& 
    -re^{2\lambda}K_2\df{}{\phi}\\
    \star & r^2\(K_5 - 2K_2\)& 0\\
    \star & 0 & r^2\sin^2\theta \(K_5 - 2K_2\)
  \end{array}\)Y_{lm}\;.
\end{align}
The expansion of the matter variables reads
\begin{align}
  \delta\eps &= \frac{\rho}{r}\,Y_{lm}\label{d_eps}\\
  \delta u_i &= -\frac{e^\nu}{r}
  \(u_1 - \frac{u_2}{r},\,u_2
  \df{}{\theta},\,u_2\df{}{\phi}\)Y_{lm}\;.
\end{align}
Of course, these expansions are to be understood as sums over all $l$
and $m$, analogous to the expansions \eqref{exp_h} and \eqref{exp_k}.
It is only through this particular choice of expansion coefficients
(e.g. in $\alpha$, $h_{rr}$ and $k_{\theta\theta}$) that we obtain a
system of equations that can be numerically integrated in a quite
straightforward and -- this is the main point -- in a stable way.

Another consequence of this decomposition is that it gives some
physical meaning to the metric perturbations $S$ and $T$. From
Eq.~\eqref{hij_polar}, we can see that $T$ represents the conformal
part of the spatial perturbation $h_{ij}$, for if we set $S=0$, the
metric is conformally flat. Obviously, $S$ itself then represents the
deviation from conformal flatness.

Finally, we should note that this decomposition is quite similar to
the one used by Allen et al.~\cite{AAKS98}. Their variables $F$ and
$S_{Allen}$ are related to ours as follows: $F = T$ and $S_{Allen} =
e^{2\nu}S$.

In order to avoid numerical problems at the origin, we have to replace
$K_1$ by the following quantity
\begin{align}\label{Kdef}
        r^2K \;:=&\; K_1 + 2r\(\df{K_2}{r} + \lam'K_2\) - K_5\;.
\end{align}
In this way, we obtain a system of five coupled evolution equations,
which are of first order in time and second order in space. There are
two equations for the metric variables $S$ and $T$ and three more for
the extrinsic curvature variables $K$, $K_2$ and $K_5$:
\begin{subequations}
  \label{FOsys}
  \begin{align}
    \label{S}
    \df{S}{t} &= K\\
    \begin{split}
      \label{K}
      \df{K}{t} &= e^{2\nu-2\lam}\bigg[
      \dff{S}{r}+ \(5\nu' - \lam'\)\df{S}{r}
      + \(4\(\nu'\)^2 + 5\frac{\nu'}{r} + 3\frac{\lam'}{r}
      - 2\frac{e^{2\lam} - 1}{r^2} - e^{2\lam}\frac{l(l+1)}{r^2}\)S\\
      &\qquad\qquad{}
      + 4\(\frac{1}{r}\(\frac{\nu'}{r}\)' + 2\(\frac{\nu'}{r}\)^2
      - \frac{\lam'\nu'}{r^2}\)T\,\bigg]
    \end{split}
  \end{align}
  \begin{align}
    \label{T}
    \df{T}{t} &= K_5\\
    \begin{split}\label{K5}
      \df{K_5}{t} &= e^{2\nu-2\lam}\bigg[
      \dff{T}{r} + \(\nu' - \lam'\)\df{T}{r}
      + \(\frac{\nu'}{r} + 3\frac{\lam'}{r}
      + 2\frac{e^{2\lam} - 1}{r^2} - e^{2\lam}\frac{l(l+1)}{r^2}\)T
      + 2\(r\nu' + r\lam' - 1\)S\,\bigg]\\
      &\qquad\qquad + 8\pi e^{2\nu}\(1 - C^2_s\)\rho
    \end{split}\\
    \label{K2}
    \df{K_2}{t} &=
    e^{2\nu-2\lam} \(r\df{S}{r} + \(2r\nu' + 1\)S
    + 2\frac{\nu'}{r}T\)\,.
  \end{align}
\end{subequations}
We could easily convert this system into a first order system both in
time and space by adding another two evolution equations for the first
derivatives $S'$ and $T'$. However, the form of the first four
equations \eqref{S} -- \eqref{K5}, which are independent of $K_2$,
suggests to rather convert them into two coupled wave equations for
$S$ and $T$
\begin{align}
  \begin{split}
    \label{waveS} \dff{S}{t} &= e^{2\nu-2\lam}\bigg[ \dff{S}{r} +
    \(5\nu' - \lam'\)\df{S}{r} + \(4\(\nu'\)^2 +
    5\frac{\nu'}{r} + 3\frac{\lam'}{r} - 2\frac{e^{2\lam} -
      1}{r^2} - e^{2\lam}\frac{l(l+1)}{r^2}\)S\\
    &{}\qquad\qquad +
    4\(\frac{1}{r}\(\frac{\nu'}{r}\)' + 2\(\frac{\nu'}{r}\)^2 -
    \frac{\lam'\nu'}{r^2}\)T\,\bigg]
  \end{split}\\
  \begin{split}
    \label{waveT}
    \dff{T}{t} &= e^{2\nu-2\lam}\bigg[
    \dff{T}{r} + \(\nu' - \lam'\)\df{T}{r}
    + \(\frac{\nu'}{r} + 3\frac{\lam'}{r}
    + 2\frac{e^{2\lam} - 1}{r^2} - e^{2\lam}\frac{l(l+1)}{r^2}\)T
    + 2\(r\nu' + r\lam' - 1\)S\,\bigg]\\
    &\qquad\qquad+ 8\pi e^{2\nu}\(1 - C^2_s\)\rho\;,
  \end{split}
\end{align}
which are equivalent to Eqs.~(14) and (15) of Allen et
al.~\cite{AAKS98}. As can be seen, the wave equation for $S$
\eqref{waveS} is totally decoupled from the fluid variable $\rho$,
which only couples to the metric perturbation $T$ in
Eq.~\eqref{waveT}. The equation for $K_2$ is only necessary in the
interior region, where it couples to the hydrodynamical equations,
which follow from energy-momentum conservation $\delta
T^{\mu\nu}_{\phantom{\mu\nu};\nu} = 0$, and are given by
\begin{subequations}
  \label{rho_system}
  \begin{align}
    \begin{split}\label{rho}
      \df{\rho}{t} &= e^{2\nu - 2\lam}\bigg[\,\df{\tilde u_1}{r} 
      - \frac{1}{r}\df{\tilde u_2}{r}
      + \(3\nu' - \lam' + \frac{1}{r}\)\tilde u_1
      - \(3\frac{\nu'}{r} - \frac{\lam'}{r}
      + e^{2\lam}\frac{l(l+1)}{r^2}\)\tilde u_2\,\bigg]\\
      &\qquad{} + \(p + \eps\)\(r\df{K_2}{r} + \(2 + r\lam'\)K_2
      - \frac{r^2}{2}K - \frac{3}{2}K_5\) + r\eps'K_2
    \end{split}\\
    \begin{split}
      \label{u1}
      \df{\tilde u_1}{t} &= C_s^2\df{\rho}{r}
      + \(\nu'\(1 + C_s^2\) + \(C_s^2\)'\)\rho
      - \half\(p + \eps\)\(r^2\df{S}{r} + \df{T}{r} + 2rS\)
    \end{split}\\
    \label{u2}
    \df{\tilde u_2}{t} &= C_s^2\rho - \half\(p + \eps\)\(r^2 S + T\)\;.
  \end{align}
\end{subequations}
Here, we have defined $\tilde u_i := \(p + \eps\)u_i$. By introducing
the enthalpy perturbation
\begin{align}\label{H_def}
  H := \frac{C^2_s}{p + \eps}\rho\;,
\end{align}
the fluid equations assume a more convenient form:
\begin{subequations}
  \label{H_system}
  \begin{align}
    \begin{split}\label{H}
      \df{H}{t} &= e^{2\nu - 2\lam}C_s^2\bigg[\df{u_1}{r} 
      + \(2\nu' - \lam'\)u_1
      + \(\frac{\lam'}{r} - 2\frac{\nu'}{r}
      - e^{2\lam}\frac{l(l+1)}{r^2}\)u_2 \bigg]
      - \nu'\(u_1 - \frac{1}{r}u_2\)\\
      &\qquad + C_s^2\(r\df{K_2}{r} + \(2 + r\lam'\)K_2
      - \frac{r^2}{2}K - \frac{3}{2}K_5\) - r\nu'K_2
    \end{split}\\
    \label{uu1}
    \df{u_1}{t} &= \df{H}{r} - \half\(r^2\df{S}{r} + \df{T}{r} + 2rS\)\\
    \label{uu2}
    \df{u_2}{t} &= H - \half\(r^2S + T\)\;.
  \end{align}
\end{subequations}
Interestingly, from Eqs.~\eqref{uu1} and \eqref{uu2}, it follows that the
coefficients $u_1$ and $u_2$ are not independent of each other but
rather are related via
\begin{align}\label{dfu2}
  u_1 &= \df{u_2}{r} + F(r)\;,
\end{align}
where $F$ is a time independent function, which has to be fixed by the
initial data. The above system \eqref{H_system}, too, can be cast into
a second order wave equation for $H$, which is equivalent to Eq.~(16)
of Allen et al.~\cite{AAKS98}~(the different signs in the terms
containing $S$ and $T$ are correct):
\begin{align}
  \begin{split}\label{waveH}
    \dff{H}{t} &= e^{2\nu - 2\lam}\bigg[\,C^2_s\dff{H}{r} 
    + \(C^2_s \(2\nu' - \lam'\) - \nu'\)\df{H}{r}
    + \(C^2_s
    \(\frac{\nu'}{r} + 4\frac{\lam'}{r} - e^{2\lam}\frac{l(l+1)}{r^2}\)
    + 2\frac{\nu'}{r} + \frac{\lam'}{r}\)H\\
    &{}\qquad + \frac{1}{2}\nu'\(C^2_s - 1\)\(r^2\df{S}{r} - \df{T}{r}\)
    + \(C^2_s\(\frac{7}{2}\frac{\nu'}{r} + \frac{\lam'}{r}
    - \frac{e^{2\lam} - 1}{r^2}\)- \frac{\nu'}{r}\(2r\nu' + \half\)\)
    \(r^2S + T\)\,\bigg]\;.
  \end{split}
\end{align}
As already mentioned in subsection B.1, we do not necessarily need
the fluid equations, for we can eliminate them by means of the
Hamiltonian constraint, which relates the fluid variable $\rho$ to the
two metric variables $S$ and $T$:
\begin{align}
  \begin{split}
    \label{HC}
    8\pi e^{2\lam}\rho &=
    - \dff{T}{r} + \lam'\df{T}{r} + r\df{S}{r}
    + \(2 - 2r\lam' +  \half e^{2\lam}l(l+1)\)S
    - \(\frac{e^{2\lam} - 1}{r^2} + 3\frac{\lam'}{r} -
    e^{2\lam}\frac{l(l+1)}{r^2}\)T\;.
  \end{split}
\end{align}
With $\rho$ substituted by the Hamiltonian constraint \eqref{HC}, the
wave equation for $T$ \eqref{waveT} in the interior now reads
\begin{align}
  \begin{split}
    \label{T_wave_in}
    \dff{T}{t} &= e^{2\nu-2\lam}C^2_s\bigg[\;
    \dff{T}{r} - r\df{S}{r} - \lam'\df{T}{r}
    +\(2r\lam' - 2 - \half e^{2\lam}l(l+1)\)S
    + \(\frac{e^{2\lam} - 1}{r^2} + 3\frac{\lam'}{r}
    - e^{2\lam}\frac{l(l+1)}{r^2}\)T\;\bigg]\\
    &\quad{} + e^{2\nu-2\lam}\bigg[\;\nu'\df{T}{r} + r\df{S}{r}
    + \(2r\nu' + \half e^{2\lam}l(l+1)\)S
    + \(\frac{\nu'}{r} + \frac{e^{2\lam} - 1}{r^2}\)T\;\bigg]\;.
  \end{split}
\end{align}
In the exterior, both $S$ and $T$ propagate with the local speed of
light $e^{\nu-\lam}$, in the interior, however, $T$ changes its
character and propagates with the local speed of sound
$e^{\nu-\lam}C_s$.

The last set of equations that is still missing are the momentum
constraints, which link the velocity perturbations to the extrinsic
curvature variables:
\begin{subequations}\label{MC}
  \begin{align}
    \begin{split}
      \label{MC1}
      16\pi e^{2\nu}\(p + \eps\)u_1 &= \df{K_2}{r} - 2\df{K_5}{r}
      + rK - \(3\nu' + 3\lam' - e^{2\lam}\frac{l(l+1)}{r}\)K_2 + 2\nu'K_5
    \end{split}\\
    \label{MC2}
    16\pi e^{2\nu}\(p + \eps\)u_2
    &= r\df{K_2}{r} - r^2K + r\(\nu' + \lam'\)K_2 - 2K_5\;.
  \end{align}
\end{subequations}
Those constraint equations have to be solved in order to obtain
physically valid initial data. We postpone the discussion on how
to construct initial to section \ref{sec:init}.

With $\rho$ being eliminated, $S$ and $T$ in the interior now become
independent variables, and the Hamiltonian constraint \eqref{HC} serves
as a definition for $\rho$. In the exterior, however, $S$ and $T$ are
not independent but have to satisfy the Hamiltonian constraint with
$\rho$ set to zero.  Unfortunately, we cannot use the Hamiltonian
constraint to further eliminate one of those variables, but it is
possible to combine $S$ and $T$ to form a new variable $Z$ (we use the
definition (20) of Allen et al.~\cite{AAKS98})
\begin{align}\label{Z}
  Z &= - \frac{2e^{2\nu}}{\Lambda l(l+1)}
  \bigg[2rT' + e^{2\lam}\(\frac{2M}{r} - 2 - l(l+1)\)T - 2r^2S\bigg]\;,
\end{align}
with
\begin{align}\label{Lambda}
  \Lambda &= l(l+1) - 2 + \frac{6M}{r}\;.
\end{align}
$Z$ then satisfies a single wave equation, the famous Zerilli equation
that was first derived in 1970 by F. Zerilli \cite{Zer70b} in the
context of black hole oscillations:
\begin{align}\label{Zeqn}
  \dff{Z}{t} &= \frac{\d^2 Z}{\d r^2_*} 
  - 2e^{2\nu}\frac{n^2(n+1)r^3 + 3n^2Mr^2 + 9nM^2r + 9M^3}
  {r^3(nr + 3M)^2}Z\;.
\end{align}
Here, we use $2n = l(l+1) - 2$, and $r_*$ is again the tortoise
coordinate defined in Eq.~\eqref{tort}. 

It is also possible to invert Eq.~\eqref{Z} and express $S$ and
$T$ in terms of $Z$ through
\begin{subequations}\label{S_T}
  \begin{align}
    T &= re^{2\nu}Z'
    + \(\half l(l+1) - \frac{6M}{r\Lambda}e^{2\nu}\)Z\\
    \begin{split}
      S &= e^{2\nu}Z''
      + \frac{M}{r^2}\(1 - \frac{6}{\Lambda}e^{2\nu}\)Z'
      + \frac{1}{r^2}\(\frac{3M}{r} - l(l+1)
      + \frac{6M}{r\Lambda}\(3 - \frac{8M}{r}\)
      - \(\frac{6M}{r\Lambda}\)^2e^{2\nu}\)Z\;.
    \end{split}
  \end{align}
\end{subequations}
Lastly, we should note that the radiated energy at infinity
can be computed from \cite{Ruoff2000}
\begin{align}
  \frac{dE}{dt} &= \frac{1}{64\pi}\sum_{l,m}\frac{(l+2)!}{(l-2)!}
  |\dot{Z}_{lm}|^2\;.
\end{align}
Since the Zerilli function contains the full polar gravitational wave
information, it should be possible to recover the original metric and
extrinsic curvature variables from it. For $S$ and $T$ we have already
given the relevant formulas in Eqs.~\eqref{S_T}. Since $K = \dot S$
and $K_5 = \dot T$, they also can computed from \eqref{S_T} with $Z$
replaced by $\dot Z$. In the exterior the momentum constraints
\eqref{MC} can be combined to give an algebraic relation for $K_2$.
Finally, the last remaining quantity $K_1$ follows from
Eq.~\eqref{Kdef}.

In the derivations of all the above equations we have made extensive
use of the computer algebra program MAPLE V in order to avoid possible
mistakes. Furthermore, we have checked that our equations and our
numerical results agree with those of Allen et al.~\cite{AAKS98}. 

\section{Boundary and junction conditions}\label{sec:bc}

There are three boundaries we have to take care of: The origin, the
surface, and the outer boundary of the numerical grid, which lies
somewhere outside the star.

At the origin $r=0$ we have to demand all variables to be regular.
From Taylor expansion around $r=0$, we can infer the analytic
behavior of the various variables. Close to the origin, $S$ and $T$,
for instance, are both proportional to $r^{l+1}$.

At the outer boundary, we impose the Sommerfeld boundary condition,
i.e.~we require the waves to be purely outgoing. If computational time
does not matter, we can even put this boundary so far away that any
contamination that enters the grid from there has not enough time to
travel to the region where we extract the signal.

The third boundary is the surface of the star at $r = R$, which is
formally defined by the vanishing of the total pressure $P$. Since the
perturbations will slightly deform the star, the perturbed surface
will be displaced by an amount $\xi^i$ with respect to the unperturbed
location at $r=R$. If the coordinates of the unperturbed surface are
denoted by $x^i_R$, the vanishing of the total pressure $P$ at the
displaced surface translates to $P(t,x^i_R +\xi^i) = 0$. From Taylor
expansion to first order we find that the (Eulerian) pressure
perturbation $\delta p$ at the surface has to obey
\begin{align}
  \delta p &= -\xi^r p'\;.
\end{align}
Unfortunately, this is not a very convenient boundary condition, since
we neither use $\delta p$ nor the displacement vector $\xi^i$ in our
set of evolution equations. Therefore we must relate this condition to
the variables we use. We will try to find a condition that gives us
the time evolution of $\delta\eps$ at the stellar surface.  The first
step is to use the relation $\delta\eps = p'/\eps'\,\delta p$, which
gives us
\begin{align}
  \df{}{t}\delta\eps &= -\eps' \df{}{t}\xi^r\;.
\end{align}
The time derivative of $\xi^r$ can then be related to the
$r$-component of the 4-velocity $u_r$ \cite{Ruoff96}:
\begin{align}
  \df{\xi^r}{t} &= e^{-2\lam}\(e^\nu\delta u_r - \beta_r\)\;.
\end{align}
After expansion in spherical harmonics, we finally obtain 
\begin{align}\label{BCrho}
  \df{}{t}\rho(t,R) &= \eps'\bigg[rK_2 + e^{2\nu-2\lam}\(u_1
  - \frac{u_2}{r}\)\bigg]_{r=R}\;.
\end{align}
The equivalent equation for the quantity $H$ as defined in \eqref{H_def}
reads
\begin{align}\label{BCH}
  \df{}{t}H(t,R) &= -\nu'\bigg[rK_2 + e^{2\nu-2\lam}\(u_1
  - \frac{u_2}{r}\)\bigg]_{r=R}\;.
\end{align}
Incidentally, this expression can be obtained directly from the
evolution equation \eqref{H} just by setting $C_s$ to zero. In the
same manner, we can obtain the boundary condition for the wave equation
\eqref{waveH}. Also, Eq.~\eqref{BCrho} follows immediately from
Eq.~\eqref{rho}, if one sets $p = \eps = p' = 0$. 

For polytropic equations of state, it is always $p = \eps = C_s = p' =
0$ at the surface of the star, hence the evolution equations
\eqref{rho} and \eqref{H} automatically lead to the right boundary
conditions. For realistic equations of state, the sound speed at the
surface should be that of iron, which is very small compared to the
sound speed inside the core, where it might reach almost the speed of
light for very relativistic stellar models. For practical purposes, in
these cases we just might as well set $C_s(r = R) = 0$. Of course,
this analysis does not hold for constant density models, but those are
not considered in this paper, anyway.

Let us now turn to the junction conditions at the surface of the star.
We will always assume that $\eps$ and $C_s$ go to zero when
approaching the stellar surface $r = R$. Following the line of
reasoning of \cite{AAKS98}, we find that $S$ is at least ${\cal C}^2$,
whereas the differentiability of $T$ depends on the value of $\rho$ at
the surface. For, if we let the subscripts $in$ and $ex$ represent the
values for the interior and the exterior, respectively, we deduce from
Eq.~\eqref{waveT} that for the second derivative of $T$ across the
surface, the following relation has to hold:
\begin{align}\label{T_inex}
  T_{in}'' - T_{ex}'' &= -8\pi e^{2\lam}\rho|_{r=R}\;.
\end{align}
From condition \eqref{BCrho}, it follows that the values of $\rho$ at
the surface depends on the value of $\eps'$ at this point. As already
discussed above, for polytropic equations of state, we have $p = \eps =
p' = 0$ at the surface. However, $\eps'$ does not necessarily vanish,
instead we have the following relation
\begin{align}\label{epssurf}
  \eps' &= -\nu'\frac{\eps^{2 - \Gamma}}{\kappa\Gamma}\;,
\end{align}
which shows that the behavior of $\eps'$ critically depends on the
value of the polytropic index $\Gamma$. We can distinguish three
different cases.  For $\Gamma < 2$, we have $\eps' \rightarrow 0$, for
$\Gamma = 2$, we have $\eps' \rightarrow$ const., whereas for $\Gamma >
2$, we have $\eps' \rightarrow -\infty$!

This is somewhat disturbing, since for the boundary condition
\eqref{BCrho} this would mean that $|\rho| \rightarrow \infty$, unless
the expression in brackets vanishes. However, this is not
automatically guaranteed! Interestingly, the boundary condition
\eqref{BCH} for $H$ is harmless for all values of $\Gamma$, since
$\nu'$ is always bounded. But, of course, in Eq.~\eqref{waveT}, we
have to use $\rho$, which we have to compute from $H$ using
\begin{align}
  \rho &= (p + \eps)\(\frac{dp}{d\eps}\)^{-1}\!H
  \;\approx\;\frac{\eps^{2 - \Gamma}}{\kappa\Gamma}H\;.
\end{align}
Again, we obtain an infinite value when $\Gamma > 2$, unless $H$
vanishes at the surface. However, as with Eq.~\eqref{BCrho}, the
boundary condition \eqref{BCH} does not guarantee the vanishing of
$H$, even if $H$ is initially set to zero.

The reason for this peculiar behavior of $\rho$ is that in the $\Gamma
> 2$ case, the Eulerian description breaks down at the surface, which
is a direct consequence of $\eps'$ becoming infinite. In this case, a
Lagrangian description of the fluid perturbations would be much more
appropriate, since for any polytropic equation of state, the
Lagrangian energy density perturbation always vanishes at the surface,
because of the vanishing of the Lagrangian pressure perturbation.

What then happens to the Eulerian density perturbation? By definition,
the Eulerian density perturbation $\delta \eps$ is the difference
between the total energy density and the background density $\eps$ at
the same location $r$, whereas the Lagrangian density perturbation
$\Delta \eps$ measures the density change in a fluid element that is
displaced by some amount $\xi^i$. It is through Taylor expansion to
linear order that we obtain the connection between the Lagrangian and
Eulerian perturbations:
\begin{align}\label{Deps}
  \Delta \eps &= \delta \eps + \xi^r\eps'\;.
\end{align}
But of course, for the $\Gamma > 2$ case this expression is ill-defined
at the surface, since the second term diverges. It is then clear that
the Eulerian perturbation $\delta\eps$ has to become infinite, too, in
order to compensate for the blow up of $\eps'$ and to yield a
vanishing Lagrangian perturbation $\Delta\eps$.

The whole discussion seems somewhat irrelevant, since we have seen
that the equations can be written in terms of metric quantities only.
However, we have to compute the second derivative of $T$, which, as
can be seen from Eq.~\eqref{T_inex}, depends on $\rho$. In the $\Gamma
> 2$ case, we therefore must have a blow up of $T''$ at the surface.
Of course, this is very troublesome for the numerical discretization,
and even for $\Gamma = 2$, we still have a discontinuity in $T''$,
which can spoil the second order convergence of the numerical
discretization scheme.

The numerical evolutions indeed confirm the above analysis. By
computing $\rho$ with the aid of the Hamiltonian constraint \eqref{HC},
we find that for polytropic stellar models with $\Gamma > 2$, $\rho$
tends to blow up at the stellar surface. For $\Gamma < 2$, we have $\rho
= 0 $ at the surface, whereas for $\Gamma = 2$, we obtain a finite
value \cite{Ruoff2000}.

\section{Constructing initial data}
\label{sec:init}
To obtain physically valid initial data, we have to solve the
constraint equations \eqref{HC} and \eqref{MC}. As is well known,
there is no unique way to do so, for those equations are
underdetermined, and therefore one always has to make some
additional assumptions about the geometry of the initial slice.

The simplest way is to choose time symmetric initial data, since in
this case the momentum constraints \eqref{MC} can be satisfied
trivially by setting all extrinsic curvature variables to zero, and we
only have to solve the Hamiltonian constraint \eqref{HC}. Of course,
time symmetric data are somewhat unphysical, since they represent a
stage of a physical system which had an arbitrary amount of incoming
radiation in the past. Nevertheless, they can give valuable insight
into the behavior of the system under consideration.

In general, all conceivable initial data for neutron star oscillations
fall into two basic categories, or combinations thereof. First, we can
assume the star being initially unperturbed. In the exterior vacuum
region, we then can arbitrarily prescribe some metric perturbations
representing a gravitational wave, which travels towards the star and
eventually excites it to oscillations. The most convenient way to do
so is to prescribe the initial Zerilli function $Z$ and its time
derivative $\dot Z$, and to compute $S$, $T$, $K$, and $K_5$ through
Eqs.~\eqref{S_T}. In this way, we have initial data that automatically
satisfy the constraints without having to solve any differential
equation.

The other possibility is to prescribe some fluid perturbations inside
the star and then use the constraints to solve for the associated
metric perturbations. As is well known, the matter distribution does
not uniquely fix the metric, since we can always superpose some
additional gravitational waves. It is therefore not clear at all what
would represent ``true'' physically realistic initial data. 

A quite common procedure is to construct conformally flat initial data
\cite{AAKLPR99}. In our case, this amounts to setting $S(t=0) = S_0 =
0$ in the Hamiltonian constraint \eqref{HC} and solving for $T_0$ for
a given initial fluid perturbation $\rho_0$. Interestingly, the thus
obtained initial data do not excite any of the spacetime modes to a
significant extent, whereas initial data with the same fluid
perturbation $\rho_0$, but with $T_0=0$ and $S_0\ne0$, can show large
$w$-mode excitations. In Fig.~1, we compare the wave signals of these
two cases for time symmetric initial data. In the case where we set
$S_0 = 0$, there is no $w$-mode signal at all, and the waveform
consists of pure fluid oscillations. In the other case, we can see a
burst of gravitational radiation, which quickly damps away through the
first $w$-mode. The final fluid ringing then coincides with the $S_0 =
0$ case.

This shows that the presence of $w$-modes is not a generic feature of
neutron star oscillations. They are only excited by a special subset
of initial data. The open question now is whether or not they will
show up in a real wave signal. Andersson and Kokkotas \cite{AK98} have
shown that by extracting the frequencies and damping times of the
$f$-mode and the first $w$-mode in a gravitational wave signal, one
can obtain important information such as mass and radius of the
neutron star. These data, in turn, can then be used to restrict the
possible equations of state. Of course, this method stands and falls
with the presence of the $w$-modes in the wave signal. But as we have
seen, it could well be that $w$-modes do not play a significant role
at all. In a subsequent paper \cite{RLP00}, we will investigate the
excitations of neutron star oscillations by means of particle
scattering, which will show that the $w$-modes can only be excited by
extremely relativistic particles.

Let us now focus on initial data which are not time symmetric. In this
case, we also have to solve the momentum constraints. Here, too, we
have some additional freedom in specifying the extrinsic curvature.
For instance we can choose Tr$K = 0$, which is equivalent to setting
\begin{align}
  K_1 &= 4 K_2 - 2K_5\;,
\end{align}
or, equivalently because of Eq.~\eqref{Kdef},
\begin{align}
  r^2K &= 2r\df{K_2}{r} + 2(r\lam' + 2)K_2 - 3K_5\;.
\end{align}
We then can substitute $K$ in Eq.~\eqref{MC} and obtain thus two coupled
differential equations for the remaining unknowns $K_2$ and $K_5$.
For given fluid perturbations $u_1$ and $u_2$, the momentum constraints
yield a unique solution.

Before presenting the numerical results, we would like to comment on
the applicability of the Lichnerowicz-York approach \cite{York79} for
the construction of initial data for neutron star perturbations, which
was used in \cite{AKLPS99}. Stated in a somewhat crude way, this
method consists in decomposing the six degrees of freedom of the
extrinsic curvature into its trace Tr$K$ (1 DOF), a transverse
trace free part $A^{TTF}_{ij}$ (2 DOFs), and a trace free longitudinal
part $A^{LTF}_{ij}$(3 DOFs). Since the momentum constraints consist
only of three equations with three source functions $j_i$, which
describe the matter distribution, only 3 DOFs are fixed, whereas the
remaining ones are freely specifiable. The simplest way is to set Tr$K
= A^{TTF}_{ij} = 0$, i.e.~the extrinsic curvature is purely
longitudinal. It can thus be written in terms of a vector potential
with three independent components, for which the momentum constraints
provide a unique solution once the sources $j_i$ and appropriate
boundary conditions have been specified.

If the extrinsic curvature is first decomposed into spherical
harmonics, this procedure still works. For the polar part, we then have
four components and two constraint equations, i.e.~only two components
can be freely specified. Here, too, by assuming Tr$K = A^{TTF}_{ij} =
0$, we can write the remaining longitudinal part in terms of a vector
potential with two polar components. Solving the constraint equations
with non-vanishing sources $j_i$ will yield nonzero values for those
component, and therefore all four polar components of the extrinsic
curvature will in general have nonzero values, either.

But this is contradiction with the Regge-Wheeler gauge, which requires
the component $\wh K_4$ to vanish. This could only be accomplished by
setting one component of the vector potential to zero. However, this
is not possible, since both components are already fixed by the
constraint equations, and in general non-vanishing.

Hence, if we want our initial data to be conform to the Regge-Wheeler
gauge, we have to have $\wh K_4 = 0$, and therefore we cannot have
both Tr$K =0$ and $A^{TTF}_{ij} = 0$. By choosing Tr$K =0$, we
already have used up all degrees of freedom to fix the extrinsic
curvature components, since we have to set $\wh K_4 = 0$, too.

\section{Numerical results for polytropic equations of state}

\subsection{Some general remarks}
Our actual goal was, by using the ADM-formalism, to obtain the
evolution equations in a hyperbolic first order form. However, due to
numerical problems at the origin, we were forced to modify those
equations in such a way that the resulting set \eqref{FOsys} is more
or less equivalent to the set of wave equations for the variables $S$,
$T$ and $H$.  Thus, from the computational point of view, there is no
point in sticking to the first order system, since the wave equations
can be numerically integrated in a much faster way. Also, we will not
explicitly integrate the fluid equations, since we can eliminate
$\rho$ in Eq.~\eqref{waveT} by means of the Hamiltonian constraint
\eqref{HC}.

For the numerical evolution, we therefore use Eqs.~\eqref{waveS}
and \eqref{T_wave_in} in the interior, whereas in the exterior we use
Eqs.~\eqref{waveS} and \eqref{waveT} with $\rho$ set to zero. Of course, in
the exterior we could try to switch from $S$ and $T$ to the Zerilli
function $Z$. We thus would have to integrate only one wave equation,
and from the resulting waveforms, we could easily read off the emitted
gravitational energy, since the radiation power is proportional to the
square of $\dot Z$.

Unfortunately, this causes numerical problems, because at the seam, we
would have to compute $Z$ and its derivatives from $S$ and $T$ and
vice versa. However, in order to obtain a correct value of $Z$, it is
crucial that $S$ and $T$ satisfy the Hamiltonian constraint. On the
other hand $S$ and $T$ satisfy the Hamiltonian constraint by
construction, at least up to discretization errors, when directly
computed from $Z$ through Eqs.~\eqref{S_T}. At the seam somewhere
outside the star, where we switch from $S$ and $T$ to $Z$, we will
always have some violation of the Hamiltonian constraint, thus the
computed Zerilli function $Z$ will not be quite correct. If we then go
back and compute $S$ and $T$ from the just obtained value of $Z$, they
will differ significantly from their original values, since now they
suddenly satisfy the Hamiltonian constraint. This mismatching at the
seam gives rise to additional reflections, which rapidly amplify
inside the star and cause the numerical code to crash after a few
dynamical time scales.

There is another point we would like to mention. When evolved with the
Zerilli equation, the amplitude of an outgoing wave remains constant
for large $r$. The same is true for $S$, however, the amplitude of $T$
grows linearly with $r$. When computing $Z$ from $S$ and $T$ by means
of formula \eqref{Z}, this growth has to cancel, what is indeed the
case when $S$ and $T$ exactly satisfy the Hamiltonian constraint. If
not, $Z$ will also start to grow for large $r$. But this is what
happens, since numerically $S$ and $T$ do not satisfy the Hamiltonian
constraint. It is predominantly the high frequency components that get
amplified the most strongly. Unfortunately, this can lead to quite
rough waveforms for $Z$, while those for $S$ and $T$ look perfectly
smooth. By increasing the resolution, this amplification will
decrease, but the resolution has to be quite high in order to obtain
accurate results.

The most practical way to obtain a quite reliable Zerilli function,
therefore, is not to completely switch to the Zerilli function in the
exterior region, but to additionally evolve $Z$ together with $S$ and
$T$. That is, close to the surface of the star we construct $Z$ from
$S$ and $T$ by means of formula \eqref{Z} and then use the Zerilli
equation \eqref{Zeqn} to independently evolve $Z$ parallel to $S$ and
$T$. Of course, this amounts to the additional computational
expenditure of evolving an extra wave equation, but we get rewarded by
obtaining much more accurate results.

To discretize the wave equations, we use the explicit second order
leap-frog scheme. If the equations are correctly implemented, the
numerical violation of the constraints should converge to zero in
second order. By monitoring the Hamiltonian constraint in the exterior,
this is indeed what we find. Of course, as mentioned above, for
polytropes with $\Gamma \ge 2$, the stellar surface can reduce the
convergence down to first order.

We should make one final note concerning the numerical treatment of
the origin $r=0$. It is well known that in radial coordinates, the
equations usually show a singular behavior close to the origin.
Moreover, Taylor expansion around the origin shows that the equations
admit two kinds of solutions, a regular and a divergent one. On
physical grounds, one usually rejects the divergent solution. For the
evolution, it is crucial that the numerical scheme preserves the
regularity condition and suppresses the divergent solution.

We have chosen the dynamical variables such that the only singular
terms are the ones proportional to $l(l+1)/r^2$. The regularity
condition requires all the perturbation variables to vanish at the
origin, however, even for $l=2$ the equations cannot be numerically
evolved with a time step size $\Delta t$ close to the maximal allowed
time step size $\Delta t_{max}$, which follows from the CFL-condition
\begin{align}
  \Delta t_{max} &= \frac{\Delta r}{c}\;,
\end{align}
where $\Delta r$ is the grid spacing in the $r$-coordinate and $c$ the
largest propagation speed, because the presence of the divergent
$l(l+1)/r^2$-terms causes the scheme to become unstable at the origin.
Only by decreasing $\Delta t$ can stability be reinforced. However,
for large values of $l$, $\Delta t$ has to be so small that it
prevents one from obtaining numerical results within a reasonable time
frame.

We therefore propose a different way that allows stable evolutions
with $\Delta t$ close to $\Delta t_{max}$ for all values of $l$.
Instead of having the boundary conditions located exactly at $r=0$, we
move it $n$ grid points to the right, i.e.~at $r_n = n\Delta r$, we impose
$S(r_n) = T(r_n) = 0$. The actual value of $n$ depends on $l$ in the
following simple way: $n = l-1$. 

The numerical experiments and a detailed stability analysis
\cite{Ruoff2000} have shown that with this little trick, we can indeed
obtain stable and second order convergent evolutions for arbitrary
values of $l$ with a Courant number of about $0.9$.

\subsection{Results}

Results concerning the evolution of various initial data have already
been presented in \cite{AAKS98}. Since the authors only focused on a
single polytropic stellar model with a central density of $\eps_0 =
3\cdot10^{15}\,$g/cm$^3$, we would like to consider three more models,
one being less relativistic and the others being more relativistic.
The physical parameters of the models are given in table 1. All
results presented are for $l=2$. Results for more stellar models and
for other values of $l$ can be found in \cite{Ruoff2000}.

As initial data, we choose a narrow time symmetric gravitational wave
pulse, centered at $r_0 = 80\,$km, where we prescribe $T$ to have a
Gaussian shape, and use the Hamiltonian constraint to compute $S$. The
numerical resolution is 500 grid points inside the star. The resulting
waveforms for the different models are shown in Fig.~2. Using a
logarithmic scale, we plot the modulus of the Zerilli function,
extracted at $r = 100\,$km.

The three waveforms clearly have quite different features. For the
least relativistic Model 1 (upper panel), there is no $w$-mode signal
at all, and the fall-off, which immediately follows the reflected wave
pulse at about $t = 0.6\,$ms, shows more a tail-like behavior. At $t =
1.1\,$ms it merges into the fluid ringing, which is dominated by the
$f$-mode. This is a somewhat unexpected result, since a direct mode
calculation reveals that for this model there do exist $w$-modes,
which should show up in the waveform. However, even the first $w$-mode
has a quite large imaginary part ($\omega$ = 0.305 + 0.24i), which
results from the model being less compact. Thus, the $w$-modes get
buried in the tail-like fall-off. The more compact and therefore more
relativistic the stellar model is, the less damped are the $w$-modes,
and they should eventually dominate over the tail-like fall-off.

This is the case for Model 2 (middle panel), which is quite close to
the stability limit. Here, instead of the tail-like fall-off, we now
can see the ring-down of the first $w$-mode ($\omega$ = 0.256 +
0.085i). Still, it is much more strongly damped than the fluid modes,
which therefore dominate the late time part.

Model 3 (lower panel), which is unstable with respect to radial
oscillations, is compact enough to allow the existence of trapped
modes. Because of their comparatively weak damping, we cannot have a
clear cut discrimination between the two ring-down phases, i.e.~at
first the $w$-modes and then the fluid modes. Instead, the signal
consists of a mixture of both the trapped and the fluid modes.
Nevertheless, the trapped modes dominate in early times, which can be
seen from the Fourier transform in the upper panel of Fig.~3. In the
lower panel, we have taken the Fourier transform at much later times,
where the most strongly damped trapped modes have damped away and the
fluid modes prevail. Still, the first two trapped modes are present in
the spectrum. 

However, there is still one feature present in the wave signal for
Model 3, which was also present in previous evolutions of the axial
modes for very compact stars, but which apparently has not been
noticed before (see e.g.~Fig.~4 of \cite{Kok97} and Fig.~5 of
\cite{KS99}).

If, in the wave signal, we determine the frequency and damping time of
the very first ring-down (from $t\approx 0.5\,$ms to about $t\approx
0.9\,$ms), we find that they do not match to any of the quasi-normal
modes associated with this particular stellar model. Instead, they
are almost identical with the complex frequency of the first
quasi-normal mode of an equal mass black hole! How is that? 

For black hole spacetimes, the quasinormal modes are derived from the
Regge-Wheeler potential in the axial case and the Zerilli potential in
the polar case. Both potentials exhibit a maximum at about $r = 3M$.
For ordinary neutron stars the surface lies usually at a radius larger
than $3M$, therefore these peaks do not exist for those stars.
However, for ultra-compact stars these peaks can lie outside.  Any
axial or polar incoming gravitational wave packet has to cross the
corresponding peak before penetrating the neutron star. While trying
to cross this peak, a large fraction of the wave will be scattered
off, which in the black hole spacetime gives rise to the quasinormal
ring-down, but since the situation is now almost identical for the
neutron star spacetime, this associated ring-down will have black-hole
characteristics.

Since inside the star, the local speed of light $e^{\nu-\mu}$ is
largely reduced, it takes a while for the remaining wave packet to be
reflected, and this is gives rise to the strong increase in the signal
at about $t=0.9\,$ms. Here, the trapped wave packet finally finds its
way out again, and it is only from thereon, that the wave signal
consists of the proper fluid and spacetime modes. (It is also from
this time when we have taken the Fourier transformation to obtain the
upper spectrum in Fig.3. If we had included the first ring-down, this
spectrum would have sat on top of a large broad peak.) For less
compact stars the local light speed inside the star is higher, and
therefore the wave packet gets reflected much earlier, making the
black-hole-like ring-down phase much shorter. We should mention that
this ring-down phenomenon for very compact stars is much more
pronounced for larger values of $l$ \cite{Ruoff2000}.

Of course, we gave a somewhat crude and heuristic explanation of this
ring-down phenomenon; we therefore postpone a more detailed study to a
future paper \cite{RK00}.

\begin{table}[t]
\caption{\label{models}List of the used polytropic stellar models
and their physical parameters.}
\begin{tabular}{ccccc}
\multicolumn{5}{c}{\rule[-2.5mm]{0mm}{7.5mm}Polytropic stellar models
($\Gamma = 2, \kappa = 100\,$km$^2$)}\\\hline
\rule[-2.5mm]{0mm}{7.5mm}
Model & $\eps_0\;[$g/cm$^3]$ & $M\;[M_\odot]$ & $R\;[$km$]$ & M/R\\
\hline
\rule[0mm]{0mm}{5mm}
1 & $1.0\!\cdot\! 10^{15}$ & 0.802 & 10.81 & 0.109\\
2 & $5.0\!\cdot\! 10^{15}$ & 1.348 & 7.787 & 0.256\\
3 & $5.0\!\cdot\! 10^{16}$ & 1.031 & 4.992 & 0.305\\
\end{tabular}
\end{table}

\section{Getting into trouble: Using realistic equations of state}
\label{sec:troubles}

So far we have used polytropic equations of state, which are quite
decent approximations to realistic equations of state as far as the
bulk features of neutron stars like mass and radius are concerned.
However, it is in particular the fluid oscillations that are very
sensitive to local changes in the equation of state, which are due to
the different behavior of the neutron star matter under varying
pressure. It is therefore important to use realistic equations of
state that take into account the underlying microphysics which
determines the state of the matter as a function of pressure and
temperature.

Realistic equations of state cannot be given in analytic terms over
the whole pressure range inside the neutron star, hence they usually
exist in tabulated form only. To solve the TOV equations in this case,
one has to interpolate between the given values in order to obtain the
stellar model with continuous functions of radius $r$. In the
following, we will make use of an equation of state called MPA
\cite{Wu91}, which yields a maximal mass model of $1.56 M_\odot$. The
non-radial oscillations modes for various stellar models have been
compiled in \cite{Pfeiffer99}.

If we try to repeat the evolution of the same initial data, but now
for a realistic stellar model with, say, $N = 200$ grid points inside
the star, we will find, after a few oscillations, an exponentially
growing mode that immediately swamps the whole evolution. It is clear
that this has to be a numerical instability, since the frequency and
$e$-folding time of this growing mode strongly depends on the chosen
resolution. In fact, by further increasing the resolution, the growing
mode starts to weaken and eventually vanishes completely. This happens
at about $N=500$ grid points inside the star, depending on the stellar
model under consideration. Hence, we have this strange fact that for
low resolutions the numerical evolution tends to be unstable, whereas
for resolutions high enough, the evolution is stable. We should stress
that this happens only with the use of realistic equations of
state. It does not happen for polytropic stellar models, even if the
resolution is extremely low.
    
Further investigations show that the origin of this instability comes
from the region close to the surface, where the sound speed has a
sharp drop. In Fig.~4 we show the profile of the sound speed $C_s =
\sqrt{dp/d\eps}$ for a stellar model using EOS MPA with a central
density of $\eps_0 = 4\!\cdot\!10^{15}$ g/cm$^3$. It can be clearly
seen that at $r \approx 8.06\,$km, there is a local minimum of the
sound speed, where it drops down to $C_s \approx 0.02$. For larger $r$
we can see a series of much smaller dips, which is an artefact of the
numerical spline interpolation between the tabulated points. But the
dip at $r \approx 8.06\,$km is physical and is related to the neutron
drip point, where the equation of state suddenly becomes very
soft. Since this occurs still in the low pressure regime, the dip is
present for any realistic equation of state.
 
It is indeed this dip in the sound speed that is responsible for the
numerical instability, for if we remove it ``by hand'', we obtain a
stable evolution! Moreover, the occurrence of the instability is
independent of the actual formulation of the equations, since the
terms that are responsible are the same in each case.  But what are
the ``bad'' terms? By simply crossing out individual terms, we find
that the culprits are those terms in the fluid equations \eqref{rho},
\eqref{H}, \eqref{waveH} or \eqref{T_wave_in} which are not multiplied
by $C_s^2$. These are the terms which remain when the sound
speed goes to zero, i.e.~these which constitute the boundary
conditions \eqref{BCrho} and \eqref{BCH}. Without these terms, the
evolution would be stable for any resolution.

The next question is then whether this instability is just a feature
of our finite differencing scheme and does not appear for a different
scheme. It turned out that the instability does indeed occur for all
discretization schemes we have tried: The central difference scheme for
the wave equations, a central difference scheme on staggered grids
and the two-step Lax-Wendroff scheme both for the first order system
of equations. Of course, these are only explicit schemes, and we do
not know whether switching to implicit schemes could be a remedy.

However, in all above cases, the instability eventually vanishes when
the resolution is increased above a certain threshold. For the central
difference schemes, this threshold is at about $N = 500$ grid points,
depending on the stellar model, whereas for the Lax-Wendroff scheme,
the required minimum resolution in order to overcome the instability
can even exceed $N = 10000$ grid points inside the star, which is
clearly unacceptable.

A more elaborate analysis of the nature of the instability can be
found in \cite{Ruoff2000}. There it is shown on the basis of a
simplified toy equation that the resolution needed in order to
overcome the instability strongly depends on the location and depth of
the dip in the sound speed. The closer to the surface and the deeper
the dip, the higher the required resolution. Since this resolution is
only needed in a very tiny region close to the surface, it would be
enough to just refine the grid in this region and use a coarser grid
outside. This could be accomplished, for instance, by fixed mesh
refinement, since this region is determined by the profile of the
sound speed, which does not change throughout the evolution. However,
we then would have to deal with the transition from the coarse grid to
the fine grid and vice versa, which might be troublesome. Another
drawback is that for each different stellar model, we would need a
different grid refinement, and it would be a matter of trial and error
to find the appropriate refinements for a stable evolution.

Yet, there is a better way out. We can try to find a new radial
coordinate $x$, which is related to the actual radial coordinate $r$
in such a way that an equidistant grid in $x$ would correspond to a
grid in $r$ that becomes automatically denser in regions where the
sound speed assumes small values. A simple relation between the grid
spacings $\Delta x$ and $\Delta r$ that has the desired property is
\begin{align}\label{DrDx}
  \Delta r(r)= C_s(r)\Delta x\;.
\end{align}
An equidistant discretization with a constant grid spacing $\Delta x$
would correspond to a coarse grid in $r$ for large values of $C_s$,
which becomes finer and finer as the sound speed $C_s$ decreases.

From the relation \eqref {DrDx}, we can immediately deduce the form of
our new coordinate $x$ as a function of $r$. By replacing the
$\Delta$'s by differentials we obtain
\begin{align}\label{transf}
  \frac{dx}{dr} &= \frac{1}{C_s(r)}
\end{align}
or 
\begin{align}\label{int_transf}
  x(r) &= \int_0^r\frac{dr'}{C_s(r')}\;.
\end{align}
As a consequence, the derivatives transform as
\begin{align}\label{diff_transf}
  \frac{d}{dx} &= C_s\frac{d}{dr}
\end{align}
and
\begin{align}
  \frac{d^2}{dx^2} &= C_s^2\frac{d^2}{dr^2} + C_sC_s'\frac{d}{dr}\;.
\end{align}
From the last relation, we see that the thus defined coordinate
transformation will transform the wave equation in such a way that the
propagation speed with respect to the $x$-coordinate will be unity
throughout the whole stellar interior.

Of course, we have to use relation \eqref{transf} with some caution,
for if $C_s = 0$ this transformation becomes singular. And this is
what happens at the stellar surface. If, for instance, the profile of
the sound speed is given in the form $C_s = C_0(1 - r/R)$, where we
have $C_s(R) = 0$, we can find an analytic expression for $x$
\begin{align}
  x(r) &= -\frac{R}{C_0}\log\(1 - \frac{r}{R}\)\;,
\end{align}
which tells us that at the surface $r = R$ we obtain $x(R) = \infty$.
In this case, the coordinate transformation seems to be quite useless,
since numerically we cannot deal with a grid that extends up to
infinity. We would have to truncate it somewhere. But from a numerical
point of view, this is not that bad, since going to infinity in the
$x$-coordinate would correspond to an infinitely fine resolution in
the $r$-coordinate at the stellar surface. But this is numerically
impossible as well, so truncating the $x$-coordinate at some point
means to define a maximal resolution in $r$ at the surface.  However,
this case usually does not happen. In all cases considered, $x(R)$ has
always a finite value.

It should be noted that the above transformation is of the same kind
as the definition of the tortoise coordinate $r_*$ in Eq.~\eqref{tort},
which also leads to wave equations with unity propagation speed
throughout the whole domain. Thus, we may call $x$ a hydrodynamical
tortoise coordinate.

Figure 5 shows the $r$-coordinate as a function of the $x$-coordinate
for three different stellar models. From the curves, it is clear that
an equidistant grid spacing in $x$ will result in an equivalent
spacing in $r$ that gets very dense towards the surface of the star,
which means that this part gets highly resolved. And this is exactly
what we need in order to overcome the instability and to obtain a
decent accuracy.

To obtain the background data on the $x$-grid, we have to solve the
TOV equations \eqref{toveqns} with respect to $x$. Since we also need
$r$ as a function of $x$, we simultaneously solve Eq.~\eqref{transf},
too. The transformed set of equations then reads:
\begin{subequations}
  \label{tovx}
  \begin{align}
    \label{tovx1}\frac{d\lam}{dx} &= C_s\(\frac{1 - e^{2\lam}}{2r}
    + 4\pi r e^{2\lam}\epsilon\)\\
    \label{tovx2}\frac{d\nu}{dx} &= C_s\(\frac{e^{2\lam} - 1}{2r}
    + 4\pi r e^{2\lam}p\)\\
    \label{tovx3}\frac{dp}{dx} &= -\frac{d\nu}{dx}(p + \eps)\\
    \label{tovx4}\frac{dr}{dx} &= C_s\;.
  \end{align}
\end{subequations}
Of course, this transformation is only defined in the stellar interior
for the fluid equations, for it is only there that the propagation
speed is the speed of sound (apart from the factor $e^{\nu-\lam}$).
If we still wanted to use Eq.~\eqref{T_wave_in}, where in the interior
$T$ plays the role of the fluid, we would have to switch from the
$x$-grid in the interior to the $r$-grid in the exterior, which is
somewhat inconvenient. It is therefore much more natural to explicitly
include the fluid equation \eqref{waveH}, which is defined in the
interior only. Thence, it is only Eq.~\eqref{waveH} which will be
transformed according to the transformation \eqref{transf}, whereas we
keep the wave equations for $S$ and $T$ as they are given in
\eqref{waveS} and \eqref{waveT}.

The transformed fluid part of the fluid equation \eqref{waveH} reads
\begin{align}
  \begin{split}
    \label{Hwave}
    \dff{H}{t} &= e^{2\nu - 2\lam}\bigg[\,\dff{H}{x}
    + \(\(2\nu_{,x} + \lam_{,x}\) - \frac{\nu_{,x}}{C_s^2}
    - \frac{C_{s,x}}{C_s}\)\df{H}{x}\\
    &{}\mbox{\hspace*{2cm}} + \(C_s\(\frac{\nu_{,x}}{r}
    + 4\frac{\lam_{,x}}{r} - C_se^{2\lam}\frac{l(l+1)}{r^2}\)
    + \frac{\lam_{,x}}{rC_s} + 2\frac{\nu_{,x}}{rC_s}\)H\,\bigg]\;.
  \end{split}
\end{align}
Here, the subscript $x$ denotes a derivative with respect to $x$. The
last missing thing is the transformation of the boundary condition
\eqref{BCH}. Unfortunately, we cannot transform Eq.~\eqref{BCH}
in a straightforward way, since at the surface it is $C_s = 0$, and
therefore the transformation of the derivative $d/dr = C_s^{-1} d/dx$
is not defined. However, the inverse transformation
\eqref{diff_transf} can make sense if we note that if $C_s = 0$ then
any derivative with respect to $x$ has to vanish. This is in
particular true for $H$ itself. Thus, at the stellar surface $r = R$,
where the sound speed $C_s$ vanishes we impose the following boundary
condition for $H$:
\begin{align}\label{bcx}
  \df{H}{x}|_{x(R)} &= 0\;.
\end{align}
This corresponds to reflection at a loose end on the $x$-grid.

Since we only transform the fluid equation and not the equations for
the metric perturbations $S$ and $T$, we have to simultaneously use
two different grids: The $x$-grid in the interior for the fluid
variable $H$, and the $r$-grid both in the interior and exterior for
the metric variables $S$ and $T$. Because of the coupling, at each
time step we have to interpolate $H$ from the $x$-grid onto the
$r$-grid in order to update equation \eqref{waveT}, and, vice versa,
both $S$ and $T$ from the $r$-grid onto the $x$-grid in order to
update equation \eqref{waveH}. This can easily be accomplished by
using spline interpolation.

In Fig.~6 we demonstrate the convergence of this new method using a
stellar model based on the realistic EOS MPA. For initial data similar
to the ones used for Fig.~2, we plot the evolution of a Gaussian wave
packet together with the evaluation of the Hamiltonian constraint for
three different spatial resolutions. By doubling the resolution, the
error in the Hamiltonian constraint should decrease by a factor of 4,
if the scheme is of second order. Conversely, the magnitude of the
error should remain roughly constant, if for each doubling of the
resolution, the error is multiplied by a factor of 4. This has been
done in Fig.~6, and one can clearly see that the magnitudes of the
thus rescaled errors remain constant for the different resolutions.
Furthermore the scheme is stable since the error does not grow with
time.

The fluid modes have their largest amplitude in the region close to
the surface. To obtain the correct mode frequencies in the power
spectrum, it is important to have high resolution close to the surface
of the star, and this is what is accomplished by our new coordinate
$x$. In Fig.~7, we show the power spectra of waveforms
obtained from runs with different resolutions. As initial data, we
chose a narrow fluid deflection at the center of the star together
with $S=0$. In this case, we suppressed any $w$-mode contribution.

It is apparent that even for the quite moderate resolution of 50 grid
points on the $x$-grid inside the star we obtain very accurate
frequencies for the first couple of fluid modes. On the $r$-grid,
however, a resolution of 200 grid points is still not enough to obtain
the same accuracy, and the peaks of the higher $p$-modes in the
spectrum are still quite far off their true values. Of course, these
results were obtained with a polytropic equation of state, since
otherwise we would not have been able to perform the evolution on the
$r$-grid at all, because of the occurring instability.

Finally, we want to demonstrate the effectiveness of our coordinate
transformation by evolving an initial perturbation for the MPA
equation of state. We choose the central energy density to be
$3\!\cdot\!10^{15}\,$g/cm$^3$, yielding a stellar mass of $M = 1.49
M_\odot$. The resolution is N = 200 grid points, which, again, would be
too low to yield a stable evolution using the $r$-coordinate. Here we
do not have any problems, the evolution is stable for {\it any} chosen
resolution. In Fig.~8, we show the power spectrum of the evolution of a
sharp initial peak in the fluid perturbation $H$, which leads to the
excitation of a multitude of fluid modes. In the interval up to
$100\,$kHz, we can find 37 modes. The direct calculation of the first
couple of modes with an eigenvalue code shows perfect agreement, which
is demonstrated in the lower panel of Fig.~8. Actually, the direct
mode calculation is much more involved and time consuming than the
evolution with a subsequent Fourier transformation. Even with a
thousand grid points inside the star, it is a matter of minutes to
obtain an accurate fluid mode spectrum for stellar models with
realistic equations of state. Of course, with this method one can only
obtain the real parts of the frequencies, the imaginary parts are too
small to be determined.

\section{Conclusions}

In this paper, we have derived, using the ADM-formalism, the
perturbation equations for non-rotating, spherically symmetric neutron
star models with barotropic equations of state. We have shown how to
choose shift and lapse and the initial data on the three dimensional
initial slice in order to obtain the Regge-Wheeler gauge. We pointed
out that not only some of the metric components but also one of the
extrinsic curvature components has to vanish in order to preserve the
Regge-Wheeler gauge throughout the evolution.

Unfortunately, the equations that directly come out after the
expansion in spherical tensor harmonics are not suited for the
numerical treatment due to problems at the origin. We therefore had to
introduce different variables that remove the singular structure of
the ``raw'' equations. Since the final system of equations in those
variables is nothing else but a system of coupled wave equations
written in first order form, it is therefore much more convenient to
use instead the second order wave equations for the numerical
evolution. Those wave equations have already been previously derived
by Allen et al.~\cite{AAKS98} in a different way. Still, our
formulation is quite useful because it facilitates the construction of
initial data in terms of metric and extrinsic curvature perturbations.

Moreover, by introducing the ADM-formalism for the derivation of the
evolution equations of non-rotating neutron stars, we have performed
the first step towards the much more involved task of deriving the
evolution equations for rotating neutron stars.  Here, the
linearization of the original Einstein equations yields quite obscure
perturbation equations, which first have to be cast into a suitable
hyperbolic form before trying to discretize them on a numerical grid.
With the ADM-equations this is a much more straightforward procedure.

We have presented results for different stellar models with polytropic
equations of state. For the least relativistic model, the wave signal
did not show any $w$-modes, although mode calculations show that they
do exist. But they are so strongly damped that the tail-like fall-off
wins over the $w$-modes. Only for more compact stellar models, the
$w$-modes are less strongly damped, and they can show up in the
waveforms. However, even for very relativistic neutron star models,
$w$-modes are not necessarily part of the waveform. In fact, their
occurrence strongly depends on the choice of initial data. For initial
data representing a gravitational wave pulse that will hit the star,
$w$-modes in general will be part of the spectrum, however, for
initial fluid perturbations, the choice of conformally flat initial
data ($S_0 = 0$) can totally suppress the presence of $w$-modes,
whereas setting $T_0=0$ and $S_0 \ne 0$, which can be regarded in some
sense as a maximal deviation from conformal flatness, shows strong
$w$-mode excitation.

For very compact stars, we found that the complex frequency of the
very first ring-down phase does not correspond to any of the proper
quasi-normal modes of the stellar model. Instead, the frequency is
almost identical with the first quasi-normal frequency of a black hole
with the same mass. We explained this phenomenon by observing that
those parts of the potential, which are responsible for the first
scattering of the impinging wave packet, lie outside the star and
therefore are the same for a black hole. Only if the part of the wave
packet that could penetrate the neutron star is reflected, the proper
ring-down with the trapped and fluid modes starts. This phenomenon,
although already present in previous evolution results
\cite{Kok97,KS99}, apparently has not been recognized before.

When switching to realistic equations of state, we are faced with a
numerical instability that is due to the sharp drop of the sound speed
at the neutron drip point, which is located quite close to the stellar
surface. Since the neutron drip point is located in the low pressure
regime, it is a feature of any realistic equation of state, and
therefore the instability should occur for all of them. The quite
peculiar fact about this instability is that it disappears when the
resolution is increased above a certain threshold.  This threshold
depends on the depth of the dip of the sound speed and is the higher
the deeper the dip is. It seems that this instability is present for
any explicit numerical discretization scheme, however, the minimum
resolution that is needed in order to suppress it, can vary
drastically for different schemes. Indeed, when using the two step
Lax-Wendroff scheme for the equations written as a first order system,
we were confronted with an almost unsurmountable resolution threshold
needed for stable evolution.

This instability does not depend on the actual formulation of the
equations and is present for any of the given formulations of this
paper. In fact, it is not only a peculiar problem of the non-radial
perturbation equations, the same instability also occurs in the radial
case \cite{Ruoff2000}; the reason being that the structure of the fluid
equations is the same in both cases.

As a remedy, we introduced the coordinate transformation
\eqref{transf}, which leads to a wave equation for the fluid with
unity propagation speed throughout the whole star. It is now possible
to evolve oscillations of realistic stellar models for any resolution.
As a further benefit, the evolution of the fluid on the $x$-grid is,
for a given number of grid points inside the star, much more accurate
than the corresponding evolution on the $r$-grid with the same number
of grid points. This is because the $x$-coordinate resolves the region
close to the stellar surface much better than the ordinary
$r$-coordinate.

This coordinate transformation could also prove useful for fast
rotating neutron stars, since the evolution equations of the fluid
will certainly have the same structure that leads to the numerical
instability in the non-rotating case. Since the rotating case is a 2D
problem, resorting to higher resolutions may become impossible, and
therefore our coordinate transformation could do the job. Maybe, this
transformation could even be of avail in nonlinear evolutions,
although it is not that straightforward to implement, since there is
no fixed background model with a static profile of the sound speed. A
possibility is to view Eq.~\eqref{transf} as a dynamic coordinate
condition that depends on the actual profile of the sound speed at
every instance of time.

\begin{acknowledgments}
  I would like to thank Pablo Laguna and Hans-Peter Nollert for
  helpful discussions and suggestions, and Kostas Kokkotas for his
  encouragement in developing this paper. I am also indebted to
  Vincent Moncrief, who kindly provided me with some of his notes.
  This work was supported by the Deutsche Forschungsgemeinschaft
  through SFB 382.
\end{acknowledgments}

\newpage
\begin{figure}[t]
\leavevmode
\epsfxsize=8.5cm
\epsfbox{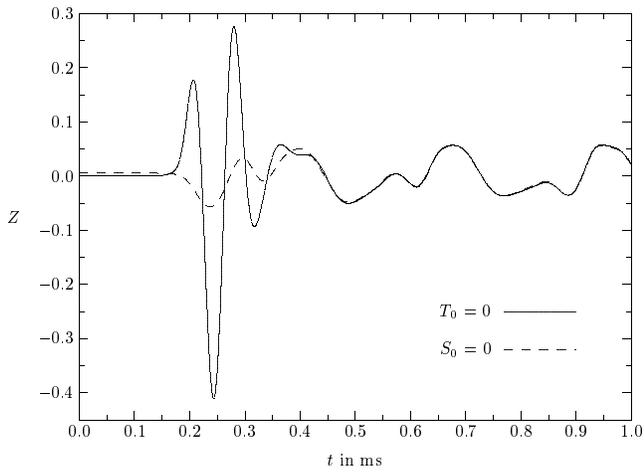}
\vspace*{5mm}
\caption{Wave signals from two different sets of initial data. The 
  signal coming from initial data with $T_0 = 0$ (solid line) shows a
  strong gravitational wave burst followed by a $w$-mode ring-down,
  whereas the signal coming from data with $S_0 = 0$ (dashed line)
  does not show any $w$-modes at all but rather consists of pure fluid
  ringing. Both signals coincide after the $w$-mode has damped away.}
\end{figure}

\begin{figure}
\leavevmode
\epsfxsize=8.5cm
\epsfbox{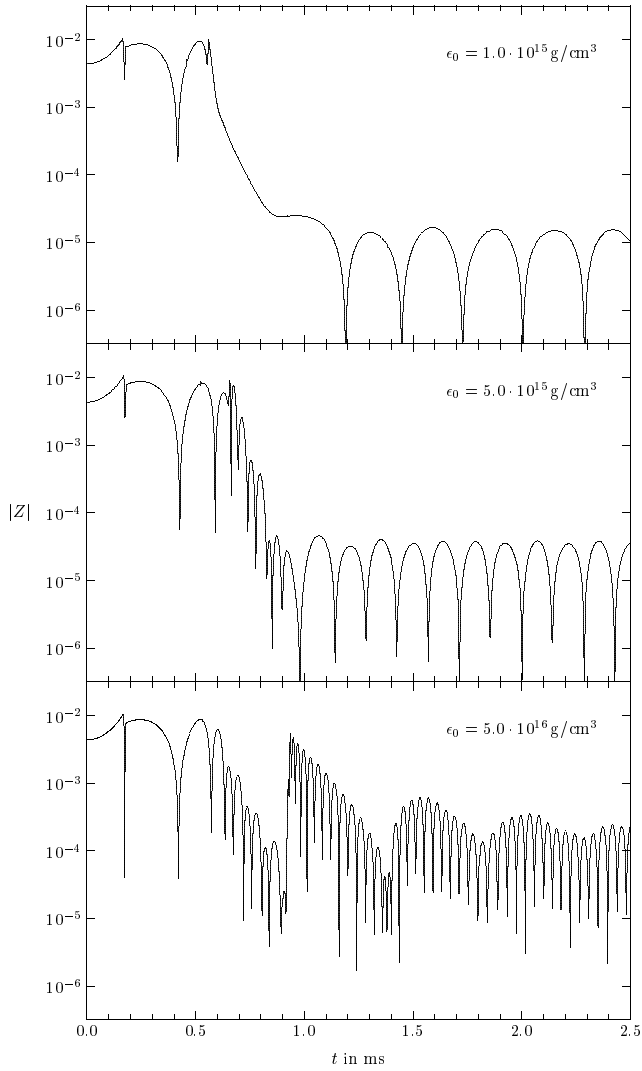}
\vspace*{5mm}
\caption{Time evolution of the Zerilli function for the three different
  polytropic stellar models M1, M2, and M3 with $l=2$. The numerical
  resolution is 500 grid points inside the star. In the least
  relativistic model (Model 1, upper panel), there are no $w$-modes
  observable.  For Model 2 (middle panel), both the $w$- and the fluid
  modes are excited. Model 3 (lower panel), which is the most compact,
  exhibits two ring-downs, one ranging from 0.6 to 0.9\,ms and
  corresponding to the first black hole quasi-normal mode, and the
  ``proper'' ring-down starting at about 1.0\,ms and consisting of the
  various trapped and fluid modes.}
\end{figure}

\begin{figure}
\leavevmode
\epsfxsize=8.5cm
\epsfbox{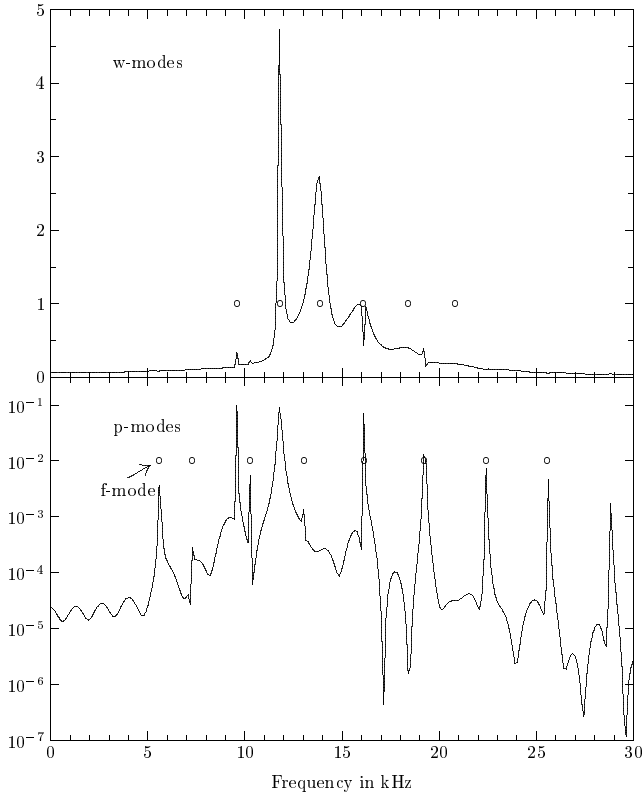}
\vspace*{5mm}
\caption{Power spectra for Model 3 with $l=2$. In the upper panel,
  the Fourier transform is taken for an early starting time ($t>
  0.9\,$ms) and shows the presence of the weakly damped trapped modes.
  In the lower panel, the Fourier transform is taken at a much later
  time, where most of the trapped modes have damped away and the
  $p$-modes prevail.  Still, the first two trapped modes are clearly
  visible.}
\end{figure}

\begin{figure}
\leavevmode
\epsfxsize=8.5cm
\epsfbox{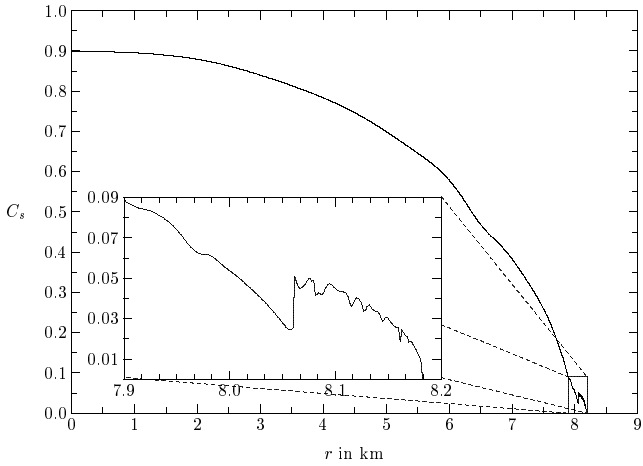}
\vspace*{5mm}
\caption{Profile of the sound speed $C_s$ inside the neutron star
  model using EOS MPA, and a section near the surface (inlet). At the
  neutron drip point around $r \approx 8.06\,$km, the sound speed
  exhibits a local minimum.}
\end{figure}

\begin{figure}
\leavevmode
\epsfxsize=8.5cm
\epsfbox{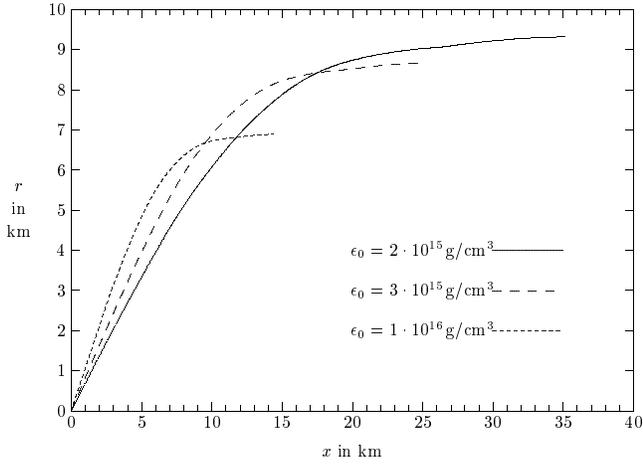}
\vspace*{5mm}
\caption{The $r$-coordinate as a function of the $x$-coordinate
  for three different stellar models. An equidistant discretization of
  $x$ corresponds to a increasingly finer resolution in $r$ as the
  stellar surface is approached.}
\end{figure}

\begin{figure}
\leavevmode
\epsfxsize=8.5cm
\epsfbox{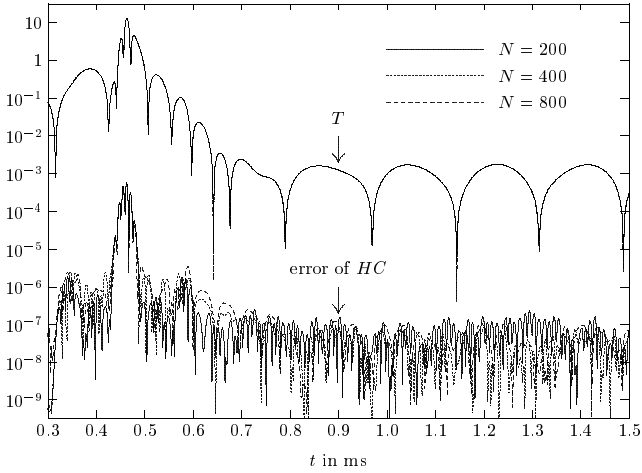}
\vspace*{5mm}
\caption{It is shown, using a realistic EOS, the evolution of the metric
  variable $T$ together with the error in the Hamiltonian constraint
  ($HC$) for the three different resolutions of $N=200$, $N=400$ and
  $N=800$ grid points inside the star. The graphs for $T$ practically
  coincide for the different resolutions. In doubling the resolution,
  the error should decrease by a factor of 4, which is compensated by
  multiplying the error by a factor of 4 for $N=400$ and by a factor of
  16 for $N=800$.  It is clearly discernible that the magnitudes of
  the scaled errors remain constant, which shows the second order
  convergence of our new numerical scheme. In addition the errors do
  not grow with time, which shows the stability.}
\end{figure}

\begin{figure}
\leavevmode
\epsfxsize=8.5cm
\epsfbox{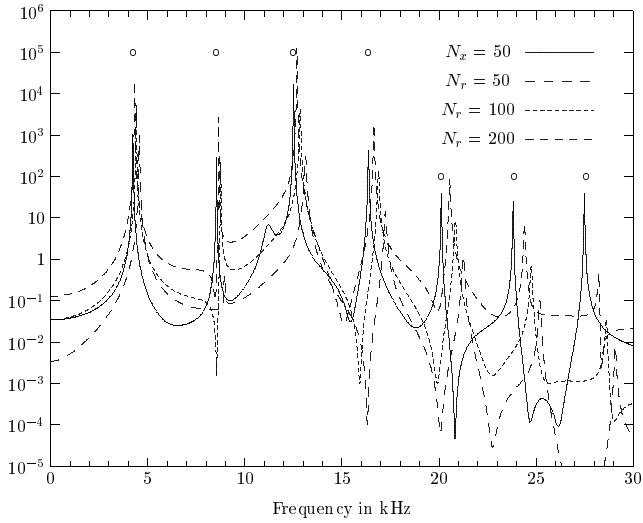}
\vspace*{5mm}
\caption{Comparison of the power spectrum obtained from an evolution
  with a resolution of $N_x = 50$ grid points on the $x$-grid with the
  spectra obtained from evolutions with different resolutions on the
  $r$-grid. Even for only 50 grid points in $x$, the peaks are closer
  to the true values than for 200 grid points in $r$.}
\end{figure}

\begin{figure}
\leavevmode
\epsfxsize=8.5cm
\epsfbox{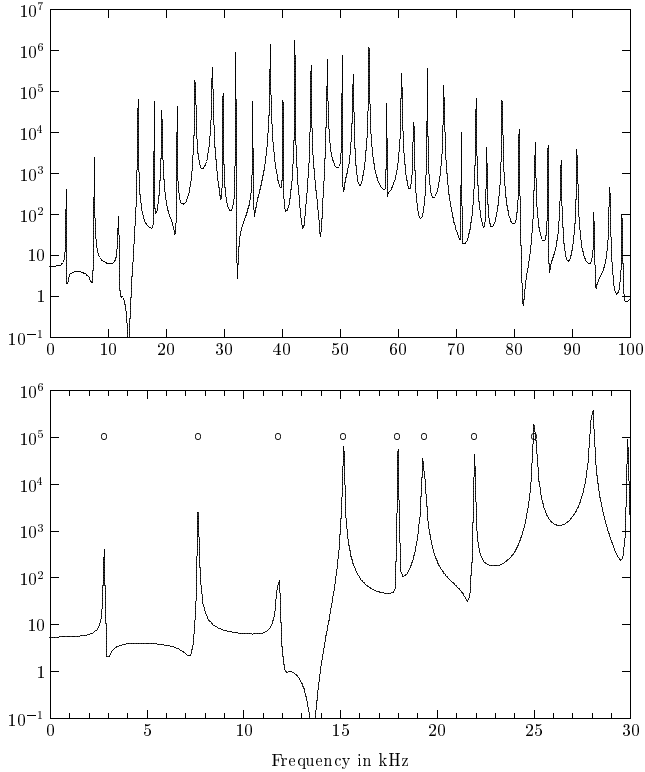}
\vspace*{5mm}
\caption{Upper panel: Power spectrum of the evolution of a sharp
  initial fluid pulse with $S_0 = 0$ on the $x$-grid. No $w$-modes are
  present, but dozens of $p$-modes are excited. In the lower panel, we
  include the modes computed by explicit mode calculation. Note that
  in contrast to polytropic equations of state, the modes do not lie
  at equidistant locations.}
\end{figure}


\begin{thebibliography}{99}

\def\pr#1#2#3{{ Phys. Rev.\ } {\bf #1}, #2 (#3)}
\def\prl#1#2#3{{ Phys. Rev. Lett.\ } {\bf #1}, #2 (#3)}
\def\prd#1#2#3{{ Phys. Rev. D} {\bf #1}, #2 (#3)}
\def\plb#1#2#3{{ Phys. Lett. B} {\bf #1}, #2 (#3)}
\def\prep#1#2#3{{ Phys. Reports} {\bf #1}, #2 (#3)}
\def\phys#1#2#3{{ Physica} {\bf #1}, #2 (#3)}
\def\jcp#1#2#3{{ J. Comput. Phys.} {\bf #1}, #2 (#3)}
\def\jmp#1#2#3{{ J. Math. Phys.} {\bf #1}, #2 (#3)}
\def\cpr#1#2#3{{ Computer Phys. Rept.} {\bf #1}, #2 (#3)}
\def\cqg#1#2#3{{ Class. Quantum Grav.} {\bf #1}, #2 (#3)}
\def\cma#1#2#3{{ Computers Math. Applic.} {\bf #1}, #2 (#3)}
\def\mc#1#2#3{{ Math. Compt.} {\bf #1}, #2 (#3)}
\def\aop#1#2#3{{ Ann. Phys.} {\bf #1}, #2 (#3)}
\def\apj#1#2#3{{ Astrophys. J.} {\bf #1}, #2 (#3)}
\def\apjs#1#2#3{{ Astrophys. J. Suppl.} {\bf #1}, #2 (#3)}
\def\acta#1#2#3{{ Acta Astronomica} {\bf #1}, #2 (#3)}
\def\sa#1#2#3{{ Sov. Astro.} {\bf #1}, #2 (#3)}
\def\sia#1#2#3{{ SIAM J. Sci. Statist. Comput.} {\bf #1}, #2 (#3)}
\def\aa#1#2#3{{ Astron. Astrophys.} {\bf #1}, #2 (#3)}
\def\mnras#1#2#3{{ Mon. Not. R. Astr. Soc.} {\bf #1}, #2 (#3)}
\def\prsla#1#2#3{{ Proc. R. Soc. London, Ser. A} {\bf #1}, #2 (#3)}
\def\ijmpc#1#2#3{{ I.J.M.P.} C {\bf #1}, #2 (#3)}
\def\nature#1#2#3{{ Nature} {\bf #1}, #2 (#3)}
\def\cmp#1#2#3{{ Commun. Math. Phys.} {\bf #1}, #2 (#3)}
\def\ptp#1#2#3{{ Prog. Theor. Phys.} {\bf #1}, #2 (#3)}
\def\grg#1#2#3{{ Gen. Rel. Grav.} {\bf #1}, #2 (#3)}
\def\cqg#1#2#3{{ Class. Quantum Grav.} {\bf #1}, #2 (#3)}
\def\ncb#1#2#3{{ Nuovo Cimento B} {\bf #1}, #2 (#3)}
\def\jsiam#1#2#3{{ J. Soc. Ind. Appl. Math.} {\bf #1}, #2 (#3)}
\def\npa#1#2#3{{ Nucl. Phys. A} {\bf #1}, #2 (#3)}
\def\rmp#1#2#3{{ Rev. Mod. Phys.} {\bf #1}, #2 (#3)}

\bibitem{Th69a} K.S. Thorne,
  \apj{158}{1}{1969}.

\bibitem{LD83} L. Lindblom and S.L. Detweiler, 
  \apjs{53}{73}{1983}.

\bibitem{Koj88} Y. Kojima,
  \ptp{79}{665}{1988}.

\bibitem{KS92}  K.D. Kokkotas and B.F. Schutz, 
  \mnras{255}{119}{1992}.

\bibitem{LNS93} M. Leins, H.-P. Nollert, and M.H. Soffel, 
  \prd{48}{3467}{1993}.

\bibitem{CF91b} S. Chandrasekhar and V. Ferrari,
  \prsla{434}{449}{1991}.

\bibitem{Kok94} K.D. Kokkotas,
  \mnras{268}{1015}{1994} + {\it Erratum}, \mnras{277}{1599}{1995}.

\bibitem{KS99} K.D. Kokkotas and B.G. Schmidt,
  {\it Quasi-normal modes of stars and black holes},
  [Article in Online Journal Living Reviews in Relativity],
  http://www.livingreviews.org/Articles/Volume2/1999-2kokkotas.

\bibitem{Nollert99} H.-P. Nollert,
  \cqg{16}{R159}{1999}.

\bibitem{FSK99} J.A. Font, N. Stergioulas, and K.D. Kokkotas,
  preprint (1999) (gr-qc/9908010).

\bibitem{AAKS98} G. Allen, N. Andersson, K.D. Kokkotas, and B.F. Schutz,
  \prd{58}{124012}{1998}

\bibitem{ADM} R. Arnowitt, S. Deser, and C.W. Misner in
  {\it Gravitation: An Introduction to Current Research}, edited
  by L. Witten (Wiley, New York, 1962), pp. 227--265.
  
\bibitem{Mon74a} V. Moncrief,
  \aop{88}{323}{1974}.

\bibitem{Mon74b} V. Moncrief,
  \aop{88}{343}{1974}.

\bibitem{CF91a} S. Chandrasekhar and V. Ferrari,
  \prsla{432}{247}{1991}.

\bibitem{IP91} J.R. Ipser and R.H. Price,
  \prd{43}{1768}{1991}.

\bibitem{Wu91} X. Wu, H. M\"uther, M. Soffel, H.Herold, and H.
  Ruder, \aa{246}{411}{1991}.
        
\bibitem{RW57} T. Regge and J.A. Wheeler,
  \pr{108}{1063}{1957}.

\bibitem{AKLPS99} N. Andersson,  K.D. Kokkotas, P. Laguna, 
  P. Papadopoulos, and M.S. Sipior,
  \prd{60}{124004}{1999}.

\bibitem{AKK96} N. Andersson, Y. Kojima, and K.D. Kokkotas,
  \apj{462}{855}{1996}.

\bibitem{Zer70b} F.J. Zerilli,
  \prl{24}{737}{1970}.
 
\bibitem{Ruoff2000} J. Ruoff, PhD Thesis, Universit\"at T\"ubingen,
  2000, electronically available at gr-qc/0010041 and 
  http://w210.ub.uni-tuebingen.de/dbt/volltexte/2000/122.
        
\bibitem{Ruoff96} J. Ruoff, Diploma Thesis, Universit\"at
  T\"ubingen, 1996.

\bibitem{AAKLPR99} G. Allen, N. Andersson,  K.D. Kokkotas, P. Laguna,
  J. Pullin, and J. Ruoff,
  \prd{60}{104021}{1999}.

\bibitem{AK98} N. Andersson and K.D. Kokkotas,
  \mnras{299}{1059}{1998}.

\bibitem{RLP00} J. Ruoff, P. Laguna, and J.Pullin,
  \prd{??}{??????}{2000}, gr-qc/0005002.

\bibitem{York79} J.W. York, in {\it Sources of Gravitational Radiation},
        edited by L.Smarr (Cambridge University Press, Cambridge, England,
        1979), p.83.
        
\bibitem{Kok97} K.D. Kokkotas, in {\it Astrophysical Sources of Gravitational Radiation},
    Eds J.A. Marck and J.P.Lasota, Cambridge University Press, 
    p.89, (1997), and gr-qc/960324.

\bibitem{RK00} J. Ruoff and K.D. Kokkotas, in preparation.

\bibitem{Pfeiffer99} E. Pfeiffer, Diploma Thesis, Universit\"at T\"ubingen, 
  1999.

\end{thebibliography}
\end{document}